\documentclass{PoS}
\usepackage{amsmath,amssymb}
\usepackage{cite}
\usepackage{bbold}
\usepackage{slashed}

\newcommand{\ii}{\mathrm{i}}

\newcommand{\dd}{\mathrm{d}}

\newcommand{\be}{\begin{equation}}
\newcommand{\ee}{\end{equation}}
\newcommand{\bea}{\begin{eqnarray}}
\newcommand{\eea}{\end{eqnarray}}
\newcommand{\e}{\mathrm{e}}
\newcommand{\del}{\partial}

\newcommand{\Tr}[1]{\:{\rm Tr}\,#1}
\def\hs#1#2{\left\langle #1\right|\left. #2\right\rangle}
\def\norm#1{\|#1\|}
\def\bra#1{\left\langle #1\right|}
\def\ket#1{\left| #1\right\rangle}
\newcommand{\gz}{\cdot}

\newtheorem{ex}{Exercise}[section]
\newcommand{\tr}[1]{\:{\rm tr}\,#1}

\newcommand{\formula}[1]{\be
#1\ee}
\newcommand{\formu}[1]{$  #1$}

\newcommand{\black}[1]{\textcolor{black}{ #1}}

\definecolor{light}{gray}{.80}
\definecolor{lightt}{gray}{.90}
\definecolor{dark}{gray}{.40}
\definecolor{GreenYellow}{cmyk}{0.15,0,0.69,0}
\definecolor{lightblue}{cmyk}{0.07,0,0,0}

\title{Noncommutative Geometry and Particle Physics}

\ShortTitle{Noncommutative Geometry and Particle Physics}

\author{Fedele Lizzi
\\
Dipartimento di Fisica ``E. Pancini'', Universit\`a di Napoli {\sl Federico II};\\
 INFN Sezione di Napoli;\\
 Departament de F\'isica Qu\`antica i Astrof\'isica
\& Institut de Ci\`encies del Cosmos, Universitat de Barcelona.\\

        E-mail: \email{fedele.lizzi@na.infn.it}}

\abstract{We review the noncommutative approach to the standard model. We start with the introduction if the mathematical concepts  necessary for the definition of noncommutative spaces, and manifold in particular. This defines the framework of spectral geometry. This is applied to the standard model of particle interaction, discussing the fermionic and bosonic spectral action. The issues relating to the calculation of the mass of the Higgs are discussed, as well as the role of neutrinos and Wick rotations. Finally, we present the possibility of solving the problem of the Higgs mass by considering a pregeometric grand symmetry.}

\FullConference{Corfu Summer Institute 2017 'School and Workshops on Elementary Particle Physics and Gravity'\\
		2-28 September 2017\\
		Corfu, Greece}

\begin{document}

There are good reasons to believe that spacetime has, at high energies, a quantum structure.
We will not dwell too much on these reasons, and just mention the semi-heuristic argument~\cite{Bronstein, Doplicher} which echoes the so called Heisenberg microscope.
The argument of the latter is the following (for a pedagogical introduction see~\cite[Sect.~1.10]{EspositoMarmoSudarshan}). Try to measure the position of a particle with an ideal microscope, this implies shining light (or some other radiation) on it, and then collecting the scattered light. A precise measurement means short wavelength. But the presence of the scales $\hbar$ of quantum mechanics, and $c$ of special relativity, imply that a short wavelength photon will have a high momentum, and the unavoidable momentum transfer in the scattering means that immediately after the measurement the information on the momentum is uncertain.
If we decide to ignore momentum, and insert $G$ and general relativity in the game, and reason at the same heuristic level,  we still have restrictions. Now the problem comes form the fact that concentrating too much energy in too small a volume creates the horizon of a black hole!
Again we have a limit to the precision of measurement.

Needless to say the Heisenberg microscope is a highly unsatisfactory way to do describe the uncertainty principle (and indeed is off by a $4\pi$ factor.) The proper way is to the formalism and interpretation of quantum mechanics where $x$ and $p$ are noncommuting operators acting on a Hilbert space, a physical state is given by a vector or a density matrix, and so on.
Unfortunately  we cannot play the same game for spacetime, as we do not have a fully fledged theory of {quantum gravity}.
Nevertheless the previous considerations suggest that a way to go is to generalise the usual geometric concepts and build a \emph{noncommutative geometry}.

We will define what we mean by noncommutative geometry as we go along, as the term may encompass different form on noncommutative spaces. We hasten to add, however, that in this case noncommutative geometry is not, or at least is not only, a spacetime with nonzero commutation relations among the position coordinates, as in~\cite{DoplicherFredenhagenRoberts}, or in countless other papers which describe variations on this theme,  their symmetries, construction of field theories on these spaces and so on. Not only it would impossible to cover all of these, but just citing the main results would make this note too long, a small portion of these can be found in~\cite{WSS} Here we will present the application of the \emph{spectral} approach to noncommutative geometry. This was pioneered by Alain Connes, and the classic reference is~\cite{connesbook}. Other good general references are~\cite{landibook,ticos,connesmarcolli,Walterbook}. Given the introductory character of the lectures we privilege review articles and textbook, whenever possible, as references.

Although the natural scale of such a generalised geometry is the Planck length, a surprising aspect is that there is an interpretation of the standard model of particle interaction (in a background gravitational field) as a very simple noncommutative geometry. Spacetime is still ``classical'', and the noncommutativity is given by a matrix algebra. The description has the possibility to be confronted with experiments and to make predictions. 

The plan of this proceeding is the following. We start with a mathematical introduction, first of topology, and then of ordinary geometry using spectral tools, we then generalise to the noncommutative case. We then proceed to the discussion of the actions (fermionic and bosonic) in noncommutative geometry. They are based on the spectral properties of a generalised Dirac operator. We apply this to the standard model of particle interaction in Sect.~\ref{se:SM}, and in Sect.~\ref{se:SMasmanif} we discuss the role of the condition for noncommutative manifolds for the standard model and beyond. In Sect.~\ref{EuclidvsLorentz} we discuss the issue of the signature in spacetime for noncommutative geometry, and in particular the Wick rotation in field theory. After the conclusions an appendix provides the discussion of the exercises present in the text.


\section{NONCOMMUTATIVE SPACES: MATHEMATICAL FOUNDATIONS. \label{math}}

In this section we will give an algebraic description of geometry, in a way which makes it suitable for the noncommutative generalization. We will start from basic topology, and then proceed to more geometrical aspects.

\subsection{From Commutative to Noncommutative Geometry.}

Geometry has been always tied closely to physics. We like to offer a quote by Galileo Galilei from the Dialogo sui massimi sistemi~\cite{Galileo}: \emph{``\`E forza confessare che il voler trattare le quistioni naturali senza geometria \`e un tentar di fare quello che \`e impossibile ad esser fatto". }\emph{(We must confess that to treat matters of nature without geometry is to attempt to do what cannot be done.)}
But what is Geometry?  To define it, we cannot think of a higher authority than Wikipedia\cite{wikipedia} \textit{Geometry (from the Ancient Greek: $\gamma\varepsilon\omega\mu\varepsilon\tau\rho\iota\alpha$  
geo- ``earth", -metron "measurement") is a branch of mathematics concerned with questions of shape, size, relative position of figures, and the properties of space}.

Geometry is an essential tool for physics, but we can make an even stronger statement. Several physical theories do not just use geometry, they \emph{are} geometry.
Think for example of analytic mechanics, which is  nothing but the symplectic geometry of phase space, or general relativity and Riemannian geometry.
The geometric view of mechanics had a crisis with the advent of the quantum world, Heisenberg's uncertainty principle made havoc of several deeply held beliefs.
All at once the notion of a point in phase space becomes untenable, and the geometry made of points need to be generalised to describe the quantum phase space. In particular the algebra of commuting functions on space becomes an algebra of quantum \emph{noncommuting} operators. Note however than one apsect did not change: in both cases the state of a physical system is a map from the element of the algebra into numbers.

\subsubsection{Commutative geometry: Hausdorff topological spaces as commutative algebras}
There are two important series of theorems, mostly due to Gelfand, Naimark and Segal, which connect ${C^*}$-algebras and Hausdorff topological spaces.

We start with the definition of ${C^*}$-algebra: ${\mathcal A}$ is an associative, normed algebra over a field (typically the complex ${\mathbb C}$), with an involution $~^*$ which satisfies the following properties:
\bea
 \norm{a+b} \leq \norm{a} + \norm{b}\ ,\ \norm{\alpha a} = |\alpha|
~\norm{a}\ ,
 \norm{ab} \leq
\norm{a}\ , \norm{b}
 \norm{a} \geq 0, \nonumber\\
\norm{a} = 0 ~\iff ~ a = 0~,\norm{a^*} = \norm a\ ,~~~ \norm{a^*a} = \norm{a}^2\ ,~~~ \forall~
a \in \mathcal A. \label{cstardef}
\eea
A $C^*$-algebra is complete in the topology given by the norm, otherwise we talk of generically of a ${*}$-algebra. 

Examples of $C^*$-algebras are ${\mathrm{Mat}(n,\mathbb C)}$, i.e.  ${n\times n}$ matrices, or the algebra of bounded or compact operators on a Hilbert space, or the algebra ${\mathcal C_0(M)}$ of continuous functions over a Hausdorff spaces ${M}$ (vanishing on the frontier in the noncompact case), with the usual pointwise multiplication. In the first case involution is Hermitean conjugacy, in the second complex conjugation of the function. One has to be careful with the norm. Obvious norms which come from a  Hilbert space structure, like ${\Tr A^\dagger A , \ A\in \mathrm{Mat}(n,\mathbb C)}$ or $\int \dd x |f(x)|^2$,  a re not good $C^*$-norms. It is easy to see that they do not satisfy the last norm property for a ${*}$-algebra in~\eqref{cstardef}. Proper $C^*$-algebra norms which satisfy~\eqref{cstardef} for the above cases are 
\bea
\|A\|^2 &=& \max_{\mbox{\scriptsize eigenvalues}} A^\dagger A ,\nonumber\\ 
\|f\|^2 &=& \sup_{x\in M} |f(x)|^2,
\eea
In particular the second case is relevant for commutative geometries.

Given a Hausdorff space ${M}$ it is always possible to define canonically a commutative ${C^*}$-algebra: just consider continuous complex valued functions. If ${M}$ is compact the algebra will be unital (i.e.\ it will contain the identity). A key result is that the inverse is also true: \emph{A commutative ${C^*}$-algebra is the algebra of continuous complex valued functions on a Hausdorff space}. 
In other words, Hausdorff spaces and ${C^*}$-algebra are in a one-to-one correspondence. The proof is constructive, given a ${C^*}$-algebra it is possible to reconstruct the points of the Hausdorff space, and their topology. For this we need the notion of state: ${\phi}$. This is a linear functional 
\be
{\phi:\mathcal A\to\mathbb C}\ee 
with the following properties:  
\begin{itemize}
\item[-]
\emph{positivity} ${\phi(a^*a)\geq 0}$ 
\item[-]  \emph{norm one} ${\|\phi\|=\sup_{\|a\|\leq 1} \phi(a)=1}$.
If the algebra is unital then it must be ${\phi(1)=1}$. 
\end{itemize}
The space of states is convex, any linear combination of states of the kind ${\left(\cos^2\lambda\right) \phi_1 + \left(\sin^2\lambda\right) \phi_2}$ is still a state for any value of ${\lambda}$. 
Some states cannot be expressed as such convex sum, they form the boundary of the set and are called pure states. 

\begin{ex} \label{ex1}
Find the states (and the pure ones) for the algebra of ${n\times n}$ matrices.
\end{ex}

\noindent For a commutative algebra the pure states coincide with the (necessarily one-dimensional) irreducible representations, as well as the set of maximal and prime ideals. In the noncommutative case these sets are different.
Gelfand and Naimark gave a prescription to reconstruct a topological space in an unique way from a commutative algebra.
The topology on the space of pure states is given by defining the limit. Given a succession of pure states ${\delta_{x_n}}$ define the limit to be
\be
{\lim_n \delta_{x_n} = \delta_x \Leftrightarrow \lim_n \delta_{x_n}(a) = \delta_x(a)\ , \forall a\in\mathcal A}.
\ee
 In other words, the states which correspond to the points are the evaluation maps, whereby the complex number associated to a function is simply the value of the function at the point.
 With the above topology the starting algebra results automatically in the algebra of continuous functions over the space of states.
 Similarly it is possible to give a topology on the set of ideals, leading to the same topology for the commutative case~\cite{FellDoran}. 

\subsubsection{Algebras as operators: the Gelfand-Naimark-Segal construction.}
In these lectures noncommutative geometry is discussed from a spectral point of view. This is a consequence of the fact that any  ${C^*}$-algebra is always representable as bounded operators on a Hilbert space. The proof is again constructive, and is called the Gelfand-Naimark-Segal (GNS) construction. We start observing that every algebra has an action on itself, where we can consider the algebra just as vector space. This will be the starting vector space we will use for the construction of the Hilbert space. The next step is the definition of an inner product with certain properties, and then, in the infinite dimensional case, it is necessary to complete the space, in the norm given by the inner product. 

Any state ${\phi}$ gives a bilinear map ${\langle a,b\rangle=\phi(a^*b)}$ with most of the properties of the inner product. 
What is missing for it to be a proper definition is the fact that there are some elements of the algebra for which ${\phi(a^*a)=0}$  even if ${a}$  is not the null vector. Those states form a (left) ideal ${\mathcal N_\phi}$. We will eliminate them  quotienting them  out.
\begin{ex} \label{2}
Prove that ${\mathcal N_\phi}$ is an ideal.  
\end{ex}
Consider now the vector space composed by the elements of the quotient, which we indicate with square bracket around any of the elements of the equivalence class: ${[a]\in\mathcal A/\mathcal N_\phi}$. The scalar product
\be
{\langle[a],[b]\rangle=\phi(a^*b)}
\ee 
 is well defined since it is easy to see that is independent on the representatives ${a}$ and ${b}$ in the equivalence classes.
We have this given a scalar product which defines a good norm. The Hilbert space is defined completing in the scalar product norm, on which ${\mathcal A}$  acts in a natural way as bounded operators.
\begin{ex}\label{3}
Perform the GNS construction for ${\mathrm{Mat}(n,\mathbb C)}$ starting from a pure state. 
\end{ex}\begin{ex}\label{4} 
Given the algebra of continuous functions on the line ${\mathcal C_0(\mathbb R)}$, consider the two states $\delta_{x_0}(a)=a(x_0)$ and ${\phi(a)=\frac1{\sqrt{\pi}}\int_{-\infty}^\infty dx\, e^{-x^2} a(x)}$. Find the Hilbert space in the two cases. 
\end{ex}

\subsubsection{Noncommutative Spaces. Morita Equivalence}
Once we have established the correspondence between Hausdorff spaces and commutative ${C^*}$-algebras, we may ask what happens for noncommutative algebras.
It is still possible to give topologies on the spaces of irreducible representations (not anymore necessarily one-dimensional), pure states and maximal or primitive ideals, but these do not coincide anymore. For us a noncomutative space will be, by extension of the concept, a noncommutative ${C^*}$-algebra. Sometimes it will possible to talk of points, at least some generalization of them, other times the concept simply does not make sense.
Think for example of the algebra of quantum operators generated by ${\hat p}$ and ${\hat q}$ of quantum mechanics. This is a deformation of the ordinary algebra of functions on phase space. Pure states are square integrable functions on ${\mathbb R}$ which are in no correspondence with the original points. Coherent states are the closest to the concept of point, but are in no way ``pointlike''. 
Consider instead the case of matrix valued functions on ${\mathbb R}$. The algebra is noncommutative,  it is nevertheless possible to discern then there is an underlying space, ${\mathbb R}$ itself, but the definitions we gave above (states, ideal, irreducible representations) not only do not coincide, but are sometimes ambiguous. We wish nevertheless to associate this algebra with a space in an algebraic way.

This last aspect  is captured by the concept of \emph{Morita Equivalence}.
We first need the concept of Hilbert Module. This is a generalization of Hilbert spaces where the field ${\mathbb C}$ is replaced by a ${C^*}$-algebra. A right Hilbert module ${\mathcal E}$ over ${\mathcal A}$ is a right module equipped with a sesqulinear form ${\mathcal E\times \mathcal E\to\mathcal A}$ linear in the first variable and with
\bea
{
 \hs{\eta_1}{\eta_2 a}_\mathcal A = \hs{\eta_1}{\eta_2}_\mathcal A a~, \ 
\hs{\eta_1}{\eta_2}_\mathcal A^* =
\hs{\eta_2}{\eta_1}_\mathcal A~,   \label{hsp1}}
\nonumber\\
{\hs{\eta}{\eta}_\mathcal A \geq 0~, ~~\hs{\eta}{\eta}_\mathcal A = 0~
\Leftrightarrow ~\eta = 0},\quad
{\forall \eta_1, \eta_2, \eta \in \mathcal E, \,a \in \mathcal A}
\eea
 Completion in the norm given by ${\|\eta\|_{\mathcal A}^2=\|\hs{\eta}{\eta}\|_{\mathcal A}}$ is assumed.
 A left Hilbert module is defined in an analogous way.
\begin{ex}\label{5}Make ${\mathcal A^N}$ into a Hilbert module, and discuss its automorphisms in the cases ${\mathcal A}$ is ${\mathrm{Mat}(n,\mathbb C)}$ or ${C_0(M)}$.
\end{ex}
Two  $C^*$-algebras ${\mathcal A}$ and ${\mathcal B}$ are said to be Morita equivalent if there exists a full right Hilbert  
${\mathcal A}$-module ${\mathcal E}$, which is at the same time a left ${\mathcal B}$-module in such a way that the structures are compatible
\be
{\hs{\eta}{\xi}_{\mathcal B} \zeta = \eta \hs{\xi}{\zeta}_{\mathcal A}~,
~~~\forall ~\eta, \xi, \zeta \in \mathcal E}. \label{compmorita}
\ee
\begin{ex}\label{6}
The algebras ${\mathbb C, \mathrm{Mat}(\mathbb C,n)}$ and and ${\mathcal K}$, compact operators on a Hilbert space, are all Morita equivalent. Find the respective bimodules.
\end{ex}
Two Morita equivalent algebras have the same space of representations, with the same topology. They describe the same noncommutative space.
The space of complex valued, or matrix valued functions, since a matrix algebra has only one representation, are Morita equivalent.
But Morita equivalence is far from being ``physical'' equivalence. There is more than topology!

\subsection{Beyond Topology: Metric Aspects.}
Let us now introduce one of the main ingredients for the Connes vision of noncommutative geometry. It joins to the algebra ${\mathcal A}$ which acts as bounded operators on a Hilbert space ${\mathcal H}$.
The third element is a self-adjoint operator ${D}$ on ${\mathcal H}$, it has compact resolvent, and therefore is bounded only for finite dimensional Hilbert spaces. It  can be seen as a generalisation of the  ${Dirac}$ operator. Indeed we will generically call it Dirac operator, even of oftentimes it will not look at all as the operator introduced by Paul Dirac ninety years ago. Together ${\mathcal A, D, \mathcal H}$ form what is called a \emph{spectral triple}.

The Dirac operator enables to describe, in purely algebraic operatorial  terms, many of  the usual structures of  geometry. Since the description is purely algebraic, it can easily be generalised to the noncommutative case, thus enabling a noncommutative geometry. The presence of ${D}$ enables for example to give a distance, and hence a metric structure, on the space of states on an algebra. The definition is
\be
{d(\phi_1,\phi_2)=\sup_{\|[D,a]\|\leq 1}\{|\phi_1(a)-\phi_2(a)|\}}.
\ee
\begin{ex}\label{7}
Take ${\mathcal A=\mathcal C(\mathbb R)}$ and  ${D=i\del_x}$. Prove that the distance among pure states gives the usual distance among points of the line ${d(x_1,x_2)=|x_1-x_2|}$ .
\end{ex}
The original Dirac operator, $\slashed\del={\gamma^\mu\del_\mu}$, ``knows'' a lot about a spin manifold. The differential structure, the spin structure,  but also the metric tensor since ${\{\gamma^\mu,\gamma^\nu\}=g^{\mu\nu}}$, 
it is the square root of the laplacian. It is therefore no surprise that it plays such an important role in pure mathematics. 
 The presence of $D$ enables an algebraic definition of one-forms and the creation of a cohomology via ${da\sim[D,a]}$. A generic one form will be ${A=\sum_i a_i[D,b_i]}$.
The construction is delicate, one has to implement ${d^2 a=0}$. To achieve this one has to quotient out some spurious ``junk'' forms, details in~\cite{landibook,ticos}. 

The dimension of an ordinary space can be obtained from the rate of growth of the eigenvalues of ${D^2}$ (Weyl). Consider the ratio of the number ${N_\omega}$ of eigenvalues smaller than a value $\omega$, divided by $\omega$ itself. Then
\be
{\lim_{\omega\to\infty} \frac{N_\omega}{\omega^{\frac d2}}} \label{dimensions}
\ee
does not diverge or vanishes for a single value of $d$, which defines the dimension, and is of course the usual Hausdorff dimension for the case of manifolds and the usual $D$. This dimension concept can be made dynamical, leading for example to the possibility of spaces which appear four dimensional at low energies, but two dimensional at high energy~\cite{LizziPinzul}.
Integrals are substituted by trace, in particular a regularised sum of eigenvalues called the Diximier trace.

A \emph{dictionary} is being built, whereby the concepts of differential are translated in a purely algebraic way.
In this way they can be generalised to the cases in which the algebra is noncommutative.
This procedure must work also in the cases which are not deformation of ordinary geometries. In some sense, since any $C^*$ can be represented as operators on a sepraable Hilbert space, any noncommutative space is in ultimate analysis an (infinite) matrix algebra, and this opens up the possibility to have finite, approximate basis. This is a huge field, which we will not pursue, for two reviews similar in spirit to what is presented here see~\cite{review1, review2}.

\subsection{Noncommutive manifolds} \label{ncmanifolds}
Manifolds play a central role in geometry, but how do we translate their differential structure? 
We need requirements which, when applied to the commutative case characterise manifolds, and which can be generalised. To do this we need to add two more ingredients to the spectral triple, they are both operators on 
${\mathcal H}$. Interestingly, both play an important role in quantum field theory. The {grading operator} $\Gamma$ with ${\Gamma^2=\mathbb 1}$ which exists only in the even case. It splits ${\mathcal H=\mathcal H_L\oplus\mathcal H_R} $. A second, antiunitary operator $J$ gives a real structure. It is connected to the Tomita-Takesaki operator. The operators are in some sense a generalization of the chirality operator $\gamma^5$ and the charge conjugation, but we prefer to use a more precise terminology.

Connes~\cite{Connesmanifold} has shown that the following seven ``axioms'' characterise manifolds in the commutative case, and generalise to the noncommutative one
\begin{enumerate}

\item{\bf Dimension.} This has been discussed above~\eqref{dimensions}.

\item{\bf Regularity.} For any $a\in{\mathcal A}$ both $a$ and
    $[D,a]$ belong to the domain of $\delta^k$ for any integer
    $k$, where $\delta$ is the derivation given by $\delta(T) =
    [|D|,T]$.

\item{\bf Finiteness.} The space $\bigcap_k {\rm Dom} (D^k)$
    is a finitely generated projective left $\mathcal A$ module.

{\item {\bf Reality.} There exist  $J$ with the commutation
    relation fixed by the number of dimensions with the property

\begin{enumerate}
\item{\emph{Commutant}.} $ [a, Jb^*J^{-1}] = 0, \forall
    a,b$

\item{\emph{First order.}} $ [[D,a], b^o = Jb^*J^{-1}] =
    0~,\forall  a,b $
\end{enumerate}
}
\item{\bf Orientation} There exists a Hochschild cycle $c$
    of degree $n$ which gives the grading $\gamma$ , This
condition gives an abstract volume form.

\item{\bf Poincar\'e duality} A Certain intersection form
    determined by $D$ and by the K-theory of $\mathcal A$ and its
    opposite is nondegenerate.

\end{enumerate}

These structures, abstract as they may seem, will be put to work in the next sections for a description of the standard model of particle interaction.

\section{ACTION!}

In this section we will apply the ideas of the previous section to quantum field theory. We will build the action for fermions and bosons, as a tool to describe the standard model of particle interactions.
We introduced the mathematical framework in a way geared for such a task. We already have (and this is of course no coincidence) the ingredients to write down the classical action of a field theory.
Before we proceed it is however necessary to state a disclaimer: we will operate in an \emph{Euclidean} and \emph{Compact}  spacetime. This is unphysical, but is a standard way to describe a field theory which solves several technical issues. 
Compactification is often  a mere technical device, to avoid infrared divergences fields are often ``defined in a box''. This may be a naive procedure however, and lately the infrared frontier is becoming relevant~\cite{Strominger, EOMAS, Scent}. Also in noncommutative filed theory the infrared plays an important role~\cite{mixing, Canfora:2015nsa,Vitale:2017nkp}.

The presence of an Euclidean spacetime is a necessity because the mathematics we built is not easily (and sometimes not at all) definable with a different signature. On the other side it is not unusual in field theory to resort to the Euclidean version,  and later the corect signature is reestablished by a Wick rotation. In our context the issue of Minkowski vs.\ Euclidean is subtle. We will get back to it in Sect.~\ref{EuclidvsLorentz}.

\begin{ex} \label{8}
Find at least two good reasons for which the construction of last lecture cannot be performed (without changes) for a noncompact and Minkowkian spacetime.
\end{ex}

The action of a quantum field theory will depend on the matter fields and on the bosons which mediate the interactions. There is a naturla choice for the matter fields. These are elements of the Hilbert space on which we represent the algebra (possibly with a reducible representation), and we will identify its elements with the fermionic matter fields of the theory. In the case of the standard model these will be quarks and leptons. The bosons instead come from the fluctuations of the Dirac operator. We define a covariant Dirac operator:
\be
D_A=D+A
\ee
where $A$ is a Hermitean one form (a potential),  i.e.\  a (generalized) one  form \formula{\sum_k a_k[D,b_k]} for generic elements of the algebra \formu{a_k,b_k\in\mathcal A}. Later we will refine the fluctuations of the operator taking the presence of real structure, and hence of antiparticles into account.
An important point is that  the algebra $\mathcal A$ determines the possible fluctuations, they are the unitary elements of the group, whose Lie algerba gives the one-form. A possible requirement of unimodularity, i.e.\ the request that the determinant of the representation of the algebra be one, may further reduce the group, by eliminating a $U(1)$ factor. Hence the elements of the spectral triple determine the symmetries of the problem.

There are two important exercises which have had a fundamental historical role~\cite{ConnesLott} and are still fundamental in understanding how and why noncommutative geometry connects to the Higgs Boson.
\begin{ex} \label{9}
Consider the toy model for which \formu{\mathcal A} is \formu{\mathbb C^2} represented as diagonal matrices on \formu{\mathcal H=\mathbb C^m\oplus \mathbb C^n}, and \formu{D_F=\begin{pmatrix} 0 & M\\M^\dagger &0\end{pmatrix}}. Find \formu{D_A}.
\end{ex}

\begin{ex} \label{10}
Consider the case for which \formu{\mathcal A} is the algebra of two copies of function on a manifold, \formu{C_0(M)\times \mathbb Z_2= C_0(M)\oplus C_0(M)}, again  represented as diagonal matrices on \formu{\mathcal H=L^2(M)\oplus L^2(M)}, and \formu{D=i \slashed \del \otimes \mathbb 1+\gamma^5\oplus D_F}.
\end{ex}
The important aspect one gather from these examples is that is that the fluctuations of such a simple example naturally give a scalar field. The curvature (the two form) than one can calculate form this case give basically the Higgs potential (the ``mexican hat'').

\subsection{Bosonic Spectral Action}

The bosonic part of the action is called the \emph{spectral action}, it was introduced originally in~\cite{Chamseddine:1996zu, Chamseddine:1996rw}. It is firmly based on the vision of geometry we introduced in Sect.~\ref{math}, and in particular on the spectrum of the (covariant) Dirac operator:
\formula{S_B=\Tr\chi\left(\frac{D_A}\Lambda\right) \label{spectralaction}}
where \formu{\chi} is a cutoff function and $\Lambda$ the energy cutoff. Often $\chi$ is taken to be the characteristic function on the interval \formu{[0,1]}. In this case the spectral action is just the number of eigenvalues of \formu{D_A} smaller than \formu\Lambda. The cutoff scale is the scale at which the standard model ceases to be effective. It may be taken to be at around the scale of unification, we will discuss this aspect in Sect.~\ref{se:renor}.

{\sl Note: Rigorously, the characteristic function is not acceptable, even if it simplifies the calculations. The problem is that it is not analytic, and this creates problems for the heat kernel expansion. Therefore one has to actually consider a smooth analytic function arbitrarily closed to the step function. Another possible choice for the cutoff, also useful for calculations, is simply a decreasing exponential.}

This action must be read in a renormalization scheme. Namely we must insert it in a path integral and read the action coming from it. It will be later expanded with heath kernel techniques.

Another possibility to regolarise the action is to use a $\zeta$-function regularization~\cite{zeta}. In this case, to obtain for the expansion also the $\lambda$ dependent terms it is necessary to understand the role of the three dimensionful  constants, namely the
cosmological constant, the Higgs vacuum expectation value, and the gravitational
constant as well as a  fundamental role of the neutrino Majorana mass term. We will see later on in Sect.~\ref{se:neutrinos} that the Neutrinos and their Majorana masses play a central role for the applications to the standard model. The issue of scales is important in this context, and the spectral action opens an interesting challenge to find a scale invariant generalization of the approach, 
where all the scales are generated dynamically. Such a formulation would allow to resolve the naturalness problem
 of the hierarchy of scales~\cite{Scaleinvariantextension}. The dynamical generation of the \emph{scale noninvariant} Einstein-Hilbert action 
is known as the "induced gravity scenario"\cite{Visser:2002ew}. There various mechanisms, which allow to generate the Higgs vacuum 
expectation value. For example it can be either triggered by the radiative corrections according to the Coleman-Weinberg mechanism 
\cite{ColemanWeinberg,GildenerWeinberg} or
it can be generated by a nontrivial dynamics of the gravitational background 
\cite{Kurkov:2016zpd}.

\subsection{Fermionic Action}

Usually the fermionic action in field theory is just:
\formula{S_F=\langle{\Psi}|D_A{\Psi}\rangle=\int \Psi^\dagger D_A \Psi}
We are in an Euclidean context, and therefore \formu{\Psi^\dagger\Psi} is the proper invariant expression under \formu{\mathrm{spin}(4)} transformations. Let us for the moment use this as fermionic action. We are ignoring the real structure, and later we will refine it.

The bosonic action is finite by construction, the fermionic
part needs to be regularised. Let us present a way to obtain the bosonic spectral action from the fermionic one based on the elimination of anomalies which occur during regularisation.
The starting point is  the fermionic action by itself. This describes a (classical) theory in which fermions move in a
fixed background.
The classical action is invariant for Weyl rescaling: 
\bea
g^{\mu\nu}&\to&\e^{2\phi} g^{\mu\nu} \nonumber\\
 \psi&\to& \e^{-\frac32\phi} \psi\nonumber\\
D_A&\to& \e^{-\frac12\phi}D\e^{-\frac12\phi} 
\eea
This is a symmetry of the classical action, not of the quantum partition function:
\be
 Z(D)=\int [\dd\psi] [\dd\psi^\dagger] e^{-S_\psi}
 \ee
 In order to make sense of this expression iy is necessary to regularise the measure, and this regularisation is not invariant.
 There is an \emph{anomaly}. 
To solve the problem we can therefore either ``correct'' the action to have an
invariant theory, or consider a theory in which the symmetry is
explicitly broken by a physical scale.
In fact we need a scale to regularise the theory anyway. 
In the spirit of noncommutative spectral geometry let us express of the partition function  formally  as a determinant:
\formula{Z(D,\mu)=\int [\dd\psi] [\dd\bar\psi] e^{-S_\psi}
=\det\left(\frac{D}{\mu}\right)}
The expression is formal because the determinant is not well defined, it diverges, and aso in this case  and we need to introduce a
cutoff. 
  The natural one in our context is a
truncation of the spectrum of the Dirac operator. This was
considered even before the appearance of the spectral action~\cite{Andrianov:1983fg,Andrianov:1983qj,Vassilevich:1987yn,Fujikawa}, it was applied to the spectral action and the standard model in~\cite{Andrianov:2010nr, Andrianov:2011jy, Andrianov:2011bc, Kurkov:2012dn}.

\def\hs#1#2{\left\langle #1\right|\left. #2\right\rangle}
\def\ketbra#1#2{\left| #1\right\rangle\left\langle #2\right|}
\def\vev#1{\left\langle #1\right\rangle}
As in the case of the bosonic spectral action, the cutoff is enforced considering only the first
\formu{N} eigenvalues of~\formu{D}.
 Consider the projector \formu{P_N=\sum_{n=0}^N
\ketbra{\lambda_n}{\lambda_n}} with \formu{\lambda_n} and \formu{
\ket{\lambda_n}} the eigenvalues and eigenvectors of \formu{D}.
The quantity  \formu{N} has become a function of the cutoff defined as
\be
N=\max n \ \mbox{such that} \ \lambda_n\leq \Lambda \label{sharpcut}
\ee
Effectively we are using \formu{N^{\mathrm{th}}}
eigenvalue as an effective cutoff. A possible choice of a different cutoff function (necessary for the bosonic part) can easily be implemented defining the determinant as a product of functions (like the $\chi$ of~\eqref{spectralaction}) of the eigenvalues. 

We can now define the regularised partition function:

\bea
Z(D,\mu)&=&\prod_{n=1}^N\frac{\lambda_n}{\mu}
=\det\left(\mathbf 1 -P_N+P_N\frac{D}{\mu}P_N\right) \nonumber\\
&=&
\det\left(\mathbf 1 -P_N+P_N\frac{D}{\Lambda}P_N\right)
\det\left(\mathbf 1 -P_N+\frac{\Lambda}{\mu}P_N\right)\nonumber\\
&=&
Z_\Lambda(D,\Lambda)
\det\left(\mathbf 1 -P_N+\frac{\Lambda}{\mu}P_N\right) \label{partfun}
\eea
The cutoff \formu{\Lambda} can be given the physical meaning of the energy in
which the effective theory has a phase transition, or at any rate an energy in
which the symmetries of the theory are fundamentally different (unification
scale). 
The possibility of a phase transition is most natural in the theory, and there are indications~\cite{Kurkov:2013kfa, Alkofer:2014raa, Besnard:2014rma} of a \emph{nongeometric} phase in that regime.
The quantity \formu{\mu} in~\eqref{partfun} is in principle different from $\Lambda$, and is a
{normalization scale}, the one which changes with the renormalization
flow
Under the change \formu{\mu\to\gamma\mu} the partition
function changes
\formula{Z(D,\mu)\to Z(D,\mu)e^{\frac1\gamma\tr P_N}}
{On the other side}
 \formula{\tr P_N=N=\tr\chi\left(\frac D\Lambda\right)=S_B(\Lambda,D)}
We found the spectral action! We obtained it with the sharp cutoff, this a consequence of our using~\eqref{sharpcut}, variations of it could give any function cutoff.
This means that we could have started considering fermions moving in a fixed background, the renormalization flow, necessary after regularization of the theory, would have given, as a backreaction, a dynamics of the bosons described by~\eqref{spectralaction}.

\section{THE STANDARD MODEL  OF PARTICLE INTERACTION \label{se:SM}}

Presently we have a very successful model which interprets the interaction of the most elementary (at present) particles. It has been confirmed by countless experiments, including the recent revelation of the Higgs boson at LHC. A particularly simple form of noncommutative geometry describes the
standard model of particle interaction, the model investigated at
CERN.

\subsection{The standard model as a Spectral Triple \label{SMtriple}}

The noncommutative geometry is particularly simple
because it is the product of an infinite dimensional commutative
algebra, times a noncommutative finite dimensional one. This kind of noncommutative space is called an \emph{almost commutative manifold.}
 Hence this algebra, being Morita equivalent of the commutative one describes a
mild generalization of the space. The infinite dimensional part is the one relative to the
four dimensional spacetime. I assume spacetime to be compact and Euclidean.

We start from the algebra, a tensor productof the continuous function over a spacetime \formu{M}  \formu{{\cal
A}=C(M)\otimes {\mathcal A}_F}, with the finite  \formu{{\cal
A}_F={\mathrm{Mat}(3,\mathbb C)\oplus{\mathbb H}\oplus\mathbb C}}, where  \formu{\mathbb H} denote quaternions, which we usually represent as two by two matrices.
The unitary elements of the algebra form a group which correspond to the correspond to the \emph{symmetries}
of the standard model: \formu{SU(3)\oplus SU(2)\oplus
U(1)}.  A unimodularity (determinant equals one) condition takes care of the
extra $U(1)$.

This algebra must be represented as operators on a
Hilbert space, a continuos infinite dimensional part
(spinors on spacetime) times a finite dimensional one: \formu{{\cal
H}={\rm sp}(\mathbb R)\otimes{\mathcal H}_F}.
 Spinors (Euclidean) come with a chirality matrix (usually called \formu{\gamma_5}), which provides the grading, and charge conjugation $J$, which associates to a spinor its (independent) hermitean conjugate. For the finite part we have a matrix \formu{\gamma} and an internal charge conjugation $J_F$ which is a matrix times complex conjugation. Together $J$ and $J_F$ provide the real structure.

This algebra must be represented as operators on a
Hilbert space, which also has a continuos infinite dimensional part
(spinors on spacetime) times a finite dimensional one: \formu{{\cal
H}={\rm sp}(M)\otimes{\mathcal H}_F}. The grading given by
\formu{\Gamma=\gamma_5\otimes \gamma} splits it into a left and right subspace:\formu{{\cal
H}_L\oplus{\mathcal H}_R}. The \formu{J} operator basically exchange the two
chiralities and conjugates, thus effectively making the algebra act
form the right.

As Hilbert space it is natural to take usual zoo of elementary particle, which transform as multiplets of standard model gauge group \formu{SU(3)\times SU(2)\times U(1)} always in the fundamental  or trivial representation of the nonabelian groups. And under \formu{U(1)} their representation is identified by the weak hypercharge \formu{Y} (which has a factor of three in the normalization)
{\begin{center}
\begin{tabular}{|c| c | c | c | c | c | c|}
\hline
 Particle  & \formu{u_R} & \formu{d_R} & \formu{\left(\begin{array}{c}u_L\\ d_L\end{array}\right)} & \formu{e_R} & \formu{\left(\begin{array}{c}\nu_L\\ e_L\end{array}\right)} & \formu{\nu_R}\\ \hline
 \formu{SU(3)} & 3&3&3&0&0&0\\ \hline
 \formu{SU(2)} & 0&0&2&0&2&0\\ \hline
  \formu{Y} & $\frac43$&$-\frac23$&$\frac13$&-2&-1&0\\ \hline
\end{tabular}
\end{center}
}
Antiparticles have hypercharge reversed,  and left-right chiralities exchanged.

Let us count the degrees of freedom:
There are 2 kind of quarks, each coming in 3 colours, and 2 leptons, this makes 8. Times 2 (eigenspaces of the chirality \formu{\gamma}). Times 2 with their antiparticles, eigenspaces of \formu{J}.
The total is 32 degrees of freedom.
Then the set of particle is repeated identically for  three generations. This brings the grand total to 96.

Note that there is some overcounting, actually a quadruplication, of states~\cite{LMMS, LMMS2, GraciaBondia:1997za, AC2M2, Gargiulo:2013bla}. On one side \formu{\mathcal H_F} contains all of the particles degrees of freedom, including for example the electrons, right and left handed, and its antipaticles, left and right positrons. On the other we take the tensor product by a Dirac spinor, whose degrees of freedom are precisely electron and positron in two chiralities. This quadruplication is actually historically called \emph{fermion doubling} because it allows fermions of mixed chirality, which happens also in lattice gauge theory where there is just a duplication.
These uncertain chirality states are unphysical, and must be projected out, they are not just overcounting. But the projection must be done only in the action, not at the level of the Hilbert space.
We will come back to this doubling/quadruplication in Sects.~\ref{beyondsta} and~\ref{EuclidvsLorentz}.

The algebra \formu{\mathcal A_F} should be represented on \formu{\mathcal H_F}. \formu{\mathbb H} must act only on right-handed particles, \formu{\mathrm{Mat}(3,\mathbb C)} only on quarks, \ldots
Moreover, we need to satisfy the various constraints of noncommutative geometry, the algebra commutes with the chirality, \formu{[a, JbJ^\dagger=0]}, \dots

We give directly the explicit expression of the representation in the basis \be
( \boldsymbol \nu_R, \boldsymbol e_R, \boldsymbol L_L,\boldsymbol u_R, \boldsymbol{d_R}, \boldsymbol Q_L, \boldsymbol \nu_R^c, \boldsymbol e_R^c, \boldsymbol L_L^c,
\boldsymbol u_R^c, \boldsymbol{d_R^c}, \boldsymbol Q_L^c ) \label{HfStruct}
\ee
where $Q_L$ corresponds to the quark doublet $(\boldsymbol u_L, \boldsymbol  d_L)$ while $L_L$ corresponds to the lepton doublet $(\boldsymbol \nu_L, \boldsymbol e_L)$, with the supercript $c$  we indicate the elements of $H_F$ which correspond to the antiparticles and by boldface characters we indicate that the elements have to replicated by three generations, for example $\boldsymbol e=(e,\mu,\tau)$ and so on. Quarks have an extra colour index, which we omit. An element of the algebra $a=(\lambda,h,m)$ with  $\lambda\in\mathbb C, h\in\mathbb H$ and $m\in\mbox{Mat}_3(\mathbb C)$ is represented by the matrix\footnote{For typographical reasons block of zeros are indicated by $\cdot$. We omit the unities, like $\otimes\mathbb 1_3$, when, for example, 
a complex number act on a quark, and likewise $\otimes\mathbb 1_2$ for doublets etc.}:
{
\begingroup
\setlength{\arraycolsep}{11.5pt}
\renewcommand{\arraystretch}{1.2}
\be
a=\left[\!\begin{array}{cccccc|cccccc}
\lambda  & \gz & \gz &  \gz & \gz & \gz & \gz & \gz & \gz & \gz & \gz & \gz \\
\gz & \lambda^* & \gz & \gz &   \gz & \gz &  \gz & \gz & \gz & \gz & \gz & \gz \\
\gz & \gz & h & \gz & \gz &  \gz & \gz & \gz & \gz &  \gz & \gz & \gz \\
 \gz & \gz & \gz & \lambda & \gz & \gz & \gz & \gz &  \gz & \gz & \gz & \gz \\
\gz &  \gz & \gz & \gz & \lambda^* & \gz & \gz & \gz & \gz & \gz & \gz & \gz \\
\gz & \gz &  \gz & \gz & \gz & h & \gz & \gz & \gz & \gz & \gz & \gz \\
\hline
\gz &  \gz & \gz & \gz & \gz & \gz & \lambda & \gz & \gz &  \gz & \gz & \gz \\
 \gz & \gz & \gz & \gz & \gz & \gz & \gz & \lambda & \gz & \gz &  \gz & \gz \\
 \gz & \gz & \gz &  \gz & \gz & \gz &\gz & \gz & \lambda & \gz & \gz & \gz \\
\gz & \gz &  \gz & \gz & \gz & \gz &  \gz & \gz & \gz & m & \gz & \gz \\
\gz & \gz & \gz & \gz & \gz & \gz & \gz & \gz & \gz & \gz & m & \gz \\
\gz & \gz & \gz & \gz & \gz & \gz & \gz & \gz & \gz & \gz & \gz & m
\end{array}\!\right].  \label{algebraconproiettori}
\ee
\endgroup
}
Note that the algebra acts in a different way on particles and antiparticle. This apparent asymmetry is solved by the presence of the real structure $J_F$ which exchanges particle with antiparticles, so that we can consider the \emph{opposite} algebra $A_F^o$ acting as $J_F A_F J_F$. Note that representation~\eqref{algebraconproiettori} is such that the actions of $A_F$ and $A_F^o$ commute. This is a nontrivial fact, and, for a generic Yang-Mills theory, there is no warranty that a representation with these properties can be found. It is true however true for the standard model, and of the Pati-Salam unified model.
 The representation can also be found in a variety of different notations in~\cite{Connesreal, Iochum:1995jr,LMMS0, DAndrea:2014ics,Walterbook}. The message I want to convey is that the fact that we have such a representation is quite ``lucky''. Very few gauge theory can have such a representation, but the standard model can.
 An important aspect is that the representations on particles and antiparticles are \emph{different}. Symmetry is restored acting on the opposite algebra \formu{J\mathcal A J^\dagger}. The real structure is therefore fundamental.

We still have to give the explicit expression for the operator \formu{D}. It
will carry the \emph{metric} information on the continuous part as
well as the internal part.
For almost commutative geometries it splits
into continuous and finite parts
\formula{D=\gamma^\mu(\del_\mu+\omega_\mu)\otimes\mathbb 1+\gamma^5\otimes
D_F}
where \formu{\omega_\mu} the spin connection. We are in a curved background. The presence of
\formu{\gamma^5}, the chirality operator for the continuous manifold
is for technical reasons.
All of the properties of the internal part are encoded in
\formu{D_F}, which is a \formu{96\times96} matrix. Inserted in the fermionic action it will eventually describe the masses of the particles, connecting each right handed fermion with its left handed counterpart, except for the  neutrino, in this case Majorana masses (connecting fermions of the same chirality= are present. In the basis~\eqref{HfStruct} the Dirac operator $D_F$ has the following explicit form:
{
\begingroup
\setlength{\arraycolsep}{11.5pt}
\renewcommand{\arraystretch}{1.2}
\be
D_F=\left[\!\begin{array}{cccccc|cccccc}
\gz & \gz &\boldsymbol  \Upsilon_\nu\!\!\! &  \gz & \gz & \gz & \boldsymbol {{\Upsilon}}^{\dagger}_R\!\!\! & \gz & \gz & \gz & \gz & \gz \\
\gz & \gz &\boldsymbol  \Upsilon_e\!\!\! & \gz &   \gz & \gz &  \gz & \gz & \gz & \gz & \gz & \gz \\
\boldsymbol \Upsilon_\nu^\dag\!\!\! & \boldsymbol \Upsilon_e^\dag\!\!\! & \gz & \gz & \gz &  \gz & \gz & \gz & \gz &  \gz & \gz & \gz \\
 \gz & \gz & \gz & \gz & \gz & \boldsymbol \Upsilon_u\!\!\! & \gz & \gz &  \gz & \gz & \gz & \gz \\
\gz &  \gz & \gz & \gz & \gz & \boldsymbol \Upsilon_d\!\!\! & \gz & \gz & \gz & \gz & \gz & \gz \\
\gz & \gz &  \gz &\boldsymbol  \Upsilon_u^\dag\!\!\! &\boldsymbol  \Upsilon_d^\dag\!\!\! & \gz & \gz & \gz & \gz & \gz & \gz & \gz \\
\hline
\boldsymbol \Upsilon_R\!\!\! &  \gz & \gz & \gz & \gz & \gz & \gz & \gz & \boldsymbol {\Upsilon}^*_\nu\!\!\! &  \gz & \gz & \gz \\
 \gz & \gz & \gz & \gz & \gz & \gz & \gz & \gz &\boldsymbol{ \Upsilon}^*_e\!\!\! & \gz &  \gz & \gz \\
\gz & \gz & \gz &  \gz & \gz & \gz &\boldsymbol  \Upsilon_\nu^t\!\!\! & \boldsymbol \Upsilon_e^t\!\!\! & \gz & \gz & \gz & \gz \\
\gz & \gz &  \gz & \gz & \gz & \gz &  \gz & \gz & \gz & \gz & \gz & 
{\boldsymbol \Upsilon}^*_u \\
\gz & \gz & \gz & \gz & \gz & \gz & \gz & \gz & \gz & \gz & \gz & {\boldsymbol \Upsilon}^*_d \\
\gz & \gz & \gz & \gz & \gz & \gz & \gz & \gz & \gz &\boldsymbol  \Upsilon_u^t\!\!\! & \boldsymbol \Upsilon_d^t\!\!\! & \gz
\end{array}\!\right].  \label{DFexplicit}
\ee
\endgroup
}
where 
\bea
\boldsymbol  \Upsilon_{\nu} &=& \hat{Y}_{u} \otimes \tilde{h}_{\nu}^{\dagger} \nonumber\\
\boldsymbol  \Upsilon_{e} &=& \hat{Y}_{d} \otimes h_{e}^{\dagger} \nonumber\\
\boldsymbol  \Upsilon_{u} &=&  \hat{y}_{u} \otimes \tilde{h}_{u}^{\dagger} \nonumber \\
\boldsymbol  \Upsilon_{d} &=&  \hat{y}_{d} \otimes {h}_d^{\dagger}\nonumber\\
\boldsymbol {\Upsilon_R^{\dagger}} &=& \hat{y}_{M}\otimes\mathrm{M}_R.  \label{choice}
\eea
In these formulas the quantities $\hat{Y}_{u}$, $\hat{Y}_{d}$, $\hat{y}_{u}$, $\hat{y}_{d}$, 
 and $ \hat{y}_{M}$ are arbitrary (dimensionless) complex 3 by 3 Yukawa matrices which act on the generation index, 
the two component columns $h_{\nu,e,u,d}$ (in the Weak isospin indexes) are chosen as follows (hereafter $v$ is an arbitrary complex constant with the dimension of mass): 
\be
h_{\nu} = \left(\begin{array}{c} v \\ 0   \end{array}\right), 
\quad h_{e} = \left(\begin{array}{c} 0 \\ v   \end{array}\right),
\quad h_{u} = \left(\begin{array}{c} v \\ 0   \end{array}\right), 
\quad h_{d} = \left(\begin{array}{c} 0 \\ v   \end{array}\right),
\ee
and the dimensionful constant  $\mathrm{M}_R$ sets the Majorana mass scale for the right handed neutrinos, needed for the sea-saw mechanism. 
The tilde in \eqref{choice} indicates charge conjugated weak isospin doublets e.g. $\tilde{h}_{\nu} = \sigma_2 {h}^*_{\nu}$, where $\sigma_2$ stands for the second Pauli matrix.

This representation is basically the only one allowed. An ambiguity for the role of the photon may be resolved considering enlarging the structure to the possibility to have nonassociative algebras, and then requiring associativity~\cite{Boyle:2016cjt}. Clifford structures suggest a different grading~\cite{DAndrea:2014ics, Dabrowski:2017hfh, Kurkov:2017wmx}.

\subsubsection*{The real structure $J_F$ }
In the basis \eqref{HfStruct} the real structure $J_F$ of the finite spectral triple reads:
{
\begingroup
\setlength{\arraycolsep}{11.5pt}
\renewcommand{\arraystretch}{1.2}
\be
J_F=\left[\!\begin{array}{cccccc|cccccc}
\gz  & \gz & \gz &  \gz & \gz & \gz & \mathbb 1 & \gz & \gz & \gz & \gz & \gz \\
\gz & \gz & \gz & \gz &   \gz & \gz &  \gz & \mathbb 1 & \gz & \gz & \gz & \gz \\
\gz & \gz & \gz & \gz & \gz &  \gz & \gz & \gz & \mathbb 1 &  \gz & \gz & \gz \\
 \gz & \gz & \gz & \gz & \gz & \gz & \gz & \gz &  \gz & \mathbb 1& \gz & \gz \\
\gz &  \gz & \gz & \gz & \gz & \gz & \gz & \gz & \gz & \gz & \mathbb 1& \gz \\
\gz & \gz &  \gz & \gz & \gz & \gz & \gz & \gz & \gz & \gz & \gz & \mathbb 1 \\
\hline
 \mathbb 1 &  \gz & \gz & \gz & \gz & \gz & \gz & \gz & \gz &  \gz & \gz & \gz \\
 \gz &  \mathbb 1 & \gz & \gz & \gz & \gz & \gz & \gz & \gz & \gz &  \gz & \gz \\
 \gz & \gz &  \mathbb 1 &  \gz & \gz & \gz &\gz & \gz & \gz & \gz & \gz & \gz \\
\gz & \gz &  \gz &  \mathbb 1 & \gz & \gz &  \gz & \gz & \gz & \gz & \gz & \gz \\
\gz & \gz & \gz & \gz &  \mathbb 1 & \gz & \gz & \gz & \gz & \gz & \gz & \gz \\
\gz & \gz & \gz & \gz & \gz &  \mathbb 1 & \gz & \gz & \gz & \gz & \gz & \gz
\end{array}\!\right] \circ cc.  \label{JF}
\ee
\endgroup
}
We emphasise that this operator is antiunitary in $H_F$.
\subsubsection*{The grading $\gamma_F$}
The last ingredient of the finite dimensional spectral triple is defined, in the basis \eqref{HfStruct}, as follows:
{
\begingroup
\setlength{\arraycolsep}{11.5pt}
\renewcommand{\arraystretch}{1.2}
\be
\gamma_F=\left[\!\begin{array}{cccccc|cccccc}
-\mathbb 1  & \gz & \gz &  \gz & \gz & \gz & \gz & \gz & \gz & \gz & \gz & \gz \\
\gz & - \mathbb 1 & \gz & \gz &   \gz & \gz &  \gz & \gz & \gz & \gz & \gz & \gz \\
\gz & \gz & \mathbb 1 & \gz & \gz &  \gz & \gz & \gz & \gz &  \gz & \gz & \gz \\
 \gz & \gz & \gz &  - \mathbb 1 & \gz & \gz & \gz & \gz &  \gz & \gz & \gz & \gz \\
\gz &  \gz & \gz & \gz & -\mathbb 1 & \gz & \gz & \gz & \gz & \gz & \gz & \gz \\
\gz & \gz &  \gz & \gz & \gz & \mathbb 1 & \gz & \gz & \gz & \gz & \gz & \gz \\
\hline
\gz &  \gz & \gz & \gz & \gz & \gz &  \mathbb 1 & \gz & \gz &  \gz & \gz & \gz \\
 \gz &  \gz & \gz & \gz & \gz & \gz & \gz & \mathbb 1 & \gz & \gz &  \gz & \gz \\
 \gz & \gz & \gz &  \gz & \gz & \gz &\gz & \gz & - \mathbb 1 & \gz & \gz & \gz \\
\gz & \gz &  \gz &  \gz & \gz & \gz &  \gz & \gz & \gz & \mathbb 1 & \gz & \gz \\
\gz & \gz & \gz & \gz &  \gz & \gz & \gz & \gz & \gz & \gz & \mathbb 1 & \gz \\
\gz & \gz & \gz & \gz & \gz & \gz & \gz & \gz & \gz & \gz & \gz & -\mathbb 1
\end{array}\!\right]  \label{gammaF}
\ee
\endgroup
}
Note that the signs of  unities on the diagonal correspond to the chiralities of the corresponding fermionic multiplets, which are equal to plus one for the left-handed particles and right-handed anti-particles and to the minus one for the right-handed particles and the left-handed antiparticles.

\subsection{The Fermionic Action and the Mass of Particles}

We must take into account the presence \formu{J} and its role for the representation of the algebra. What is important is that if we want to have the proper representations of the gauge groups acting on particles and antiparticles in the proper way, then we must emply the ``right action'' which is given by \formu{J}.
 This has an effect on the fluctuations of \formu{D} which must take into account this, and therefore become
\formula{D_A=D+A+JAJ^\dagger}
for a generic one-form A.
Again I will not give the details~\cite{Connesreal, Iochum:1995jr, Walterbook}, but it is easy to  imagine that commuting with \formu{D_F} with the element of \formu{\mathrm{Mat}(3,\mathbb C)} give gluons, commuting with the quaternions gives the \formu{W} and with \formu{U(1)} gives the \formu{B} field.

Here we come to the central point of the noncomutative geometry approach to the standard model:  there are also the fluctuations in the internal space, which, as in the example we have seen, give a \emph{scalar field} with all the characteristics of the Higgs boson! This a very nice aspect of the model. The Higgs boson is built just in the same way of all other bosons of the standard model, and is on a par with them. While the former are built commuting with the spacetime part, the Higgs emerges from commuting with the internal part, and this is the reason it is a scalar and not a vector.

Taking \formu{J} into account the fermionic action must be written as
\formula{S_F=\langle J\Psi|D_A\Psi\rangle \label{SFwithJ}}
Inserting in the appropriate places of the matrix \formu{D_F} the masses (Yukawa couplings) of the particles, gives the proper mass terms.
 Let me stress that the calculation it a bit lengthy, but totally straightforward, basically commuting matrices of rank at most 3. 

\subsection{ Bosonic Action}
If the function $\chi$ in~\eqref{spectralaction} is exactly a step function then
\formula{S_B=\tr\chi\left(\frac{D_A}{\Lambda}\right)=\tr\chi\left(\frac{D_A^2}{\Lambda^2}\right)}
otherwise the functions in the two traces are slightly different. We will assume in the following the step function. We want to express this action in terms of the potential one-form, its curvature, the spin connection, Riemann tensors etc. The standard technique used in that of the \emph{heath kernel}. An excellent review of this technique can be found in~\cite{manual}.

Technically the bosonic spectral action is a sum of residues and can
be expanded in a power series in terms of $\Lambda^{-1}$ as
\be
S_B=\sum_n f_n\, a_n(D^2/\Lambda^2)
\ee
where the $f_n$ are the momenta of $\chi$
\begin{eqnarray}
f_0&=&\int_0^\infty \dd x\, x  \chi(x)\nonumber\\
f_2&=&\int_0^\infty \dd x\,   \chi(x)\nonumber\\
f_{2n+4}&=&(-1)^n \del^n_x \chi(x)\bigg|_{x=0} \ \ n\geq 0
\end{eqnarray}
the $a_n$ are the Seeley-de Witt coefficients which vanish for $n$
odd. For $D^2$ of the form
\be
D^2=-(g^{\mu\nu}\del_\mu\del_\nu\mathbb 1+\alpha^\mu\del_\mu+\beta)
\ee

Defining (in term of a generalised spin connection containing also the gauge
fields)
\begin{eqnarray}
\omega_\mu&=&\frac12 g_{\mu\nu}\left(\alpha^\nu+g^{\sigma\rho} \Gamma^\nu_{\sigma\rho}\mathbb 1\right)\nonumber\\
\Omega_{\mu\nu}&=&\del_\mu\omega_\nu-\del_\nu\omega_\mu+[\omega_\mu,\omega_\nu]\nonumber\\
E&=&\beta-g^{\mu\nu}\left(\del_\mu\omega_\nu+\omega_\mu\omega_\nu-\Gamma^\rho_{\mu\nu}\omega_\rho\right)
\end{eqnarray}
then

\begin{eqnarray}
a_0&=&\frac{\Lambda^4}{16\pi^2}\int\dd x^4 \sqrt{g}
\tr\mathbb 1_F\nonumber\\
a_2&=&\frac{\Lambda^2}{16\pi^2}\int\dd x^4 \sqrt{g}
\tr\left(-\frac R6+E\right)\nonumber\\
a_4&=&\frac{1}{16\pi^2}\frac{1}{360}\int\dd x^4 \sqrt{g}
\tr(-12\nabla^\mu\nabla_\mu R +5R^2-2R_{\mu\nu}R^{\mu\nu}\nonumber\\
&&+2R_{\mu\nu\sigma\rho}R^{\mu\nu\sigma\rho}-60RE+180E^2+60\nabla^\mu\nabla_\mu
E+30\Omega_{\mu\nu}\Omega^{\mu\nu})
\end{eqnarray}
$\tr$ is the trace over the inner indices of the finite algebra
$\mathcal A_F$ and in $\Omega$ and $E$ are contained the gauge
degrees of freedom including the gauge stress energy tensors and the
Higgs, which is given by the inner fluctuations of $D$. The other coefficients in the expansion are suppressed by powers of $\Lambda$, they can be calculated, and could play a role in allowing other multiparticle vertices, they have been studied in~\cite{a6}.

With these tools we now turn to calculations in the standard model.

\subsection{Renormalization and the mass of the Higgs \label{se:renor}}

In this section we will start to ``get numbers'', i.e.\ we will insert the actions we have seen before in a renormalization flow, fixing the boundary conditions via experimental values, and show how to get, for example, the mass of the (Englert, Brout, Guralnick, Hagen, Kibble) Higgs boson. We will also discuss the meaning and validity of this calculation, and wheter we can deem it to be a \emph{prediction}.

Let us remind the ingredients we have gathered so far:

\begin{itemize}

\item An algebra is a product of a continuous infinite dimensional part and a discrete finite dimensional noncommutative part. The algebra is the product \formu{C_0(M)\times (\mathrm{Mat}(3,\mathbb C)\oplus \mathbb H \oplus \mathbb C)}.

\item A representation of the algebra on a Hilbert space containing the known fermions. The representation is aymmetric, we present it in~\eqref{algebraconproiettori}.
\item A generalised Dirac operator which has information on the curved background of the continuous Riemannian part as well as the masses of the fermions.

\item A chirality and a charge conjugation operator.

\item An action based on the spectrum of the operators, which we expand in series.

\end{itemize}

What is left is just cranking a machine, which as I said is a straightforward but lengthy exercise. The calculation is described, in the form of the spectral action I am presenting here, in~\cite{AC2M2}, where, in Sect.~4.1 the Lagrangian of the standard model coupled with background gravity is presented in its full glory and gory details. I will not rewrite here the 14 lines equation. The was not to ``discover'' this Lagrangian, which was of course known, but to find which noncommutative geometry was able to describe it.

\begin{ex}\label{gut}
Perform a similar construction for a Grand Unified Theory
\end{ex} 

Here comes however a little, welcome, surprise.  The various constants which appear are all fixed by the entries of \formu{D_F}, which are the Yukawa couplings of the fermions and their mixing.
It turns out that the parameters of the Higgs are dependent on these, and are not independent, as in the standard model.
{Do we have a prediction of the Higgs?}

We have written a Lagrangian, but it  is still a classical one. We 
have to quantize it, and implement on it the renormalization programme\footnote{The Lagrangian is written coupling the standard model to gravity. I will  not quantize the gravitational field (which would give a nonrenormalizable theory)}, to perform this we  choose as background and ordinary Minkowki space.
Renormalisation means that all ``constant'' quantities in the action become a functions of the energy: \emph{running coupling constants}.
The running is given by the \formu{\beta}  function, solution of an ordinary differential
equation, calculated perturbatively to first
(occasionally second, rarely third) order in \formu{\hbar} by the \formu{n}-point
amplitude (loop expansion).l At this stage we use traditional quantum field theory techniques.

First think one has to decide is at which energy one write
the big Lagrangian. This will give a boundary condition. It turns out that the fact that the top quark mass is much higher with respect to the other particles (\formu{\sim 170~\mathrm{GeV}} vs.\ \formu{\sim 4~\mathrm{GeV}} for the bottom) the Higgs parameters as well as the flow are dominated by this value. 

\begin{figure}[htp]
 \includegraphics[scale=1.8]{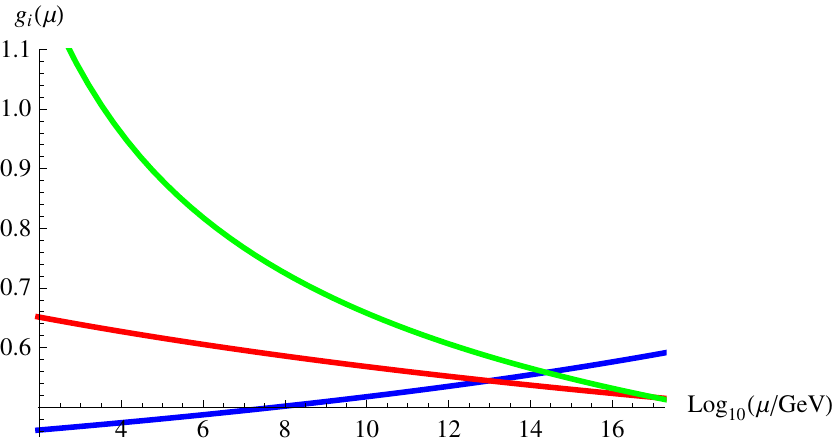}
 \caption{\sl The running of the three gauge coupling constants. Where $g_i$ correspond to the coupling to U(1), SU(2) and SU(3). For U(1) a $\frac35$ normalization factor has been taken into account.}  \label{fig1}
\end{figure}

One boundary condition for the running of the gauge coupling constants can be given by the fact that in the
model obtained ``cranking the machine'' the strength of the fundamental interaction is equal.
Experimentally is known that, \emph{if there are no other particles appearing at higher energy}, then the three coupling constant are \emph{almost}
equal at one scale, as can be seen from Fig.~\ref{fig1}. The presence of other particles may change the running, leading for example to a single pole~\cite{Majani,ULP}, and also gravitational effects may play a role at high energy, as has been investigated in~\cite{Devastato:2013wza}.

The Dirac operator $D_F$ in~\eqref{DFexplicit} contains all Yukawa couplings, data relative to the fermions, but no information on the Higgs mass namely vacuum expectation value and quartic coupling coefficient. These turn out to be functions of the parameters in $D_F$ in a way which dominated by the top quark mass (which is much larger than all other masses). This means that we can \emph{caculate} the Higgs mass (which in turn is a function of quartic coupling and vacuum expectation value.

We have a ``prediction'' for the Higgs mass.
The prediction is \formu{175.1+5.8-7.2 \mathrm{GeV}}. And unfortunately this number is wrong\ldots
The actual experimental value is \formu{ 125.09\pm 32 \mathrm{GeV}}. 

Let me make a little sociological comment. In these lectures, in particular in this section, I have been using with very little mathematical rigour. This may disconcert, or even horrify mathematicians. Yet, in physics we have a very stringent rigour: experimental verification. We let ourselves do anything we want with a theory, and in the end we judge it by its predictions.
Now it depends how you consider this theory. If you take it as a mature fully formed theory then the result is wrong. That's it! Throw away the theory.
 Alternatively, if you take it (as I do) as a tool to investigate the standard model starting from first principles, then it is remarkable that a theory based on pure mathematical result gets reasonable numbers, and  take  the measurement of the Higgs as a reason to understand in which direction one has to improve on the theory. And I hasten to add that it possible to reconcile it with experiment.

The prediction depends on the boundary condition, and we were forced to use one for which the coupling constants of the three interactions were equal at a scale. This is true only approximately, but the picture I showed is based on a running which starts from low energy (where we have data) and extrapolates to high energy.
{ But if one changes the field content, then the runnings of \emph{all} quantities change}
{ It is known that in some supersymmetric theories the presence of the supersymmetric partners alters the running and could lead to unification.}

\subsection{ Right handed neutrinos \label{se:neutrinos}}

{ Enter right handed neutrinos. Thy are the most recent addition to the particle zoo. We have indirect evidence of their existence from the fact that neutrino oscillate between different flavours. This is possible only for massive particles.}
{There is not (yet) a direct measurement of the mass of left handed neutrinos, but oscillation data shows that it must be very small, \formu{\lesssim 0.12~\mathrm{eV}}. To compare, considering all particles in increasing mass order,  the following mass is that of the electron at~~\formu{0.511~\mathrm{MeV}}. More than six orders of magnitude heavier!}
{Although origin of the numerical values of all masses in the standard model is rather mysterious, the reason for such a low value is doubly so. And it suggests that the neutrino masses might have a different origin from the other masses.}

{Presently the most popular hypothesis is to use the so called \emph{see-saw mechanism}, i.e.\ give a high \emph{Majorana} mass to the right handed neutrino, in addition to the usual Dirac mass. Here by ``high'' we mean of the order of the unification energy, the scale at which the standard model would change its order parameter.}
{ Usually particles have Dirac masses, which connect spinor with different chiralities: \formu{\psi_L m \psi_R}}
{But also \emph{Majorana} masses are possible in the Lagrangian, this connect same chirality spinors \formu{\psi_R m_M \psi_R} or \formu{\psi_L m'_M \psi_L}}. I have actually anticipated myself by already considering Majorana masses $\mathrm M_R$ for neutrinos in~\eqref{DFexplicit}.
There is a
mass matrix comprising both kind of masses, with the Majorana masses on the diagonal and the Dirac mass off diagonal.
In the presence of generations the matrices are not just two by two, but the idea is the same.
The eigenstate of mass are obtained diagonalizing this matrix. The \emph{see-saw} mechanism assumes a matrix in which there is a Majorana mass for the right handed neutrino of the order of the unification scale, and a ``normal'' smaller Dirac mass. One of the eigenvalues of this matrix  is \formu{\sim m_M} while the other is \formu{m/m_M^2\ll m}.
The constraints given by noncommutative geometry are quite stringent. Nevertheless, it turns out that the slot corresponding to a Majorana mass matrix for right handed neutrino is allowed. One might think that the addition of this  term in $D_F$ would change the running. After all we have another large scale in the model, and this may change the running of the \formu{\beta} functions, and cause on one side the unification of the three interactions, and on the other a different value for the Higgs mass. Moreover the extra term may give rise to extra bosonic fields.

Surprisingly the  explicit calculation shows that the extra term in \formu{D_F} \black{commutes} with the algebra, and therefore no extra boson, no different value for the Higgs.
This suggested Chamseddine and Connes~\cite{resilience} to consider the entry in \formu{D_F} to be an independent \emph{field}.
{This extra field has some of the properties of the Higgs, and it couples to it, changing the running.}
{Such a field had already appeared in the literature when it was realised that a relatively light Higgs at \formu{125~\mathrm{GeV}} creates a dangerous \black{instability}. At some energy, intermediate between the present scale and the unification scale, the coefficient of the quartic term in the Higgs potential becomes \black{negative}, turning the Mexican hat into a potential not bounded from below.}
{The problem is that in noncommutative geometry all bosons (so far are in one-forms which are obtained by commuting \formu{D} with elements of the algebra. In this way we obtained \formu{W}, \formu{Z}, photons gluons and the Higgs. Putting this field by hand is at best unpleasant.}
{Moreover the addition of an extra boson, with its mass and coupling parameter lowers the predictive power of the model, with the addition of the extra parameter one can only state that the model is compatible with data.}
{We should probably ask more to the model! And go back to its roots.}

\section{THE STANDARD MODEL AND BEYOND AS A NONCOMMUTATIVE MANIFOLD \label{se:SMasmanif}}

{Since the conditions for a spectral triple to describe a manifold have been cast algebraically in Sect.~\eqref{ncmanifolds}, we can see which noncommutative finite dimensional \formu{C^*} algebras satisfy the conditions. And I remind you that a finite dimensional \formu{C^*} is necessarily a sum of matrices over the reals, complex or quaternions}
{This is again a straightforward exercise of finite dimensional linear algebra, and it has been performed in~\cite{AC2M2}. It is necessary to use need use all of the five elements of the triple. The result is that the finite part of the spectral triple must have a well defined form:}
\formula{\mathbb M(\mathbb H)_a \oplus \mathbb M(\mathbb C)_{2a}\label{manifoldalgebra}}
for \formu{a} integer.
{A noncommutative manifold is the direct sum of matrices of quaternions (which in turn can be represented as \formu{2\times 2} matrices) and matrices of complex number \emph{of the same size}.}
{This algebra has to be represented on a Hilbert space of dimension \formu{n=2(2a)^2} (up to generation replicas).}
The gauge group is made of the unitary operators, and the symmetry will be ``broken'', thus reducing the gauge group.
{There is not much freedom in the game, and we must verify if there is way to obtain the algebra of the standard model.}
{ This algebra acts on a finite Hilbert space of dimension \formu{2(2a)^2}. 
For a non trivial grading it must be  \formu{a\geq 2}.} This means that the first nontrivial case we have to examine is:
\formula{
{\mathcal A}_F= 
\mathbb{M}_{2}(\mathbb{H})\oplus\mathbb{M}_{4}(\mathbb{C})}
Acting on a Hilbert space of dimension \formu{2(2\cdot2)^2=32}, the
dimension of \formu{{\mathcal H}_F} for one generation!

{The grading condition \formu{[a,\Gamma]=0}
reduces the algebra to the left-right algebra}
\formula{
\mathcal{A}_{LR}=\mathbb{H}_L\oplus\mathbb{H}_R\oplus\mathbb{M}_{4}(\mathbb{C}) \label{ALR}}
The unimodular elements of this algebra form the group \formu{SU(4)\times SU(2) \times SU(2)}. The
order one condition reduces further the algebra to ${\mathcal A}_{sm} $, i.e.\ the standard model algebra.
The group \formu{SU(4)\times SU(2) \times SU(2)} has been introduced long ago~\cite{PatiSalam} and goes under the name of Pati-Salam model. Is one of the first example of a Grand Unified Theory. There usual $SU(3)$ of colour is enlarge to $SU(4)$, with the lepton number playing the role of fourth colour, and the ypercharge $U(1)$ is enlarge to a right handed $SU(2)$, to make the model left-right symmetric.
In order to obtain the standard model another breaking mechanism is needed, at an higher scale. A field,  analog to the Higgs, which we will call \formu{\sigma}, is necessary. This field appears in  \formu D in the position corresponding to a particular form of the neutrino mass (Majorana). It turns out that precisely in that spot (and others) it is possible to put a nonzero value!

Unfortunately, as we said,  cranking of the machine does not produce a contribution to the one form, the extra term commutes with the algebra. One possibility is to include it by hand~\cite{resilience}. This is unpleasant, since so far all of standard model bosons are the result of the fluctuations of $D_F$, which is a matrix of constant numbers, and does not contain fields.
Nevertheless,  performing again the running of the physical quantitates with this field does change the Higgs mass, making it compatible with the experimental value
Physics is therefore telling us that into his framework right handed neutrinos, and Majorana masses are crucial.

\subsection{Beyond Standard \label{beyondsta}}

It would be nice to avoid this insertion ad hoc. Three possible solutions have been offered.

\begin{itemize}

\item {One possibility is to enlarge the Hilbert space introducing new fermions and new $U(1)$ interactions~\cite{Stephan:2009te}. } 

\item  Consider enlarging the symmetry, still within the framework of Sect.~\ref{ncmanifolds}, and keeping the same Hilbert space. 

\item  It is possible to consider the violation of one (or more) of the conditions in Sect.~\ref{ncmanifolds}. In particular in~\cite{Chamseddine:2013kza, Chamseddine:2013rta} the order one condition is violated.

\end{itemize}

{The two latter solutions allow the introduction of a new field \formu{\sigma} which not only fixes the mass of the Higgs making it compatible with 126~GeV, but also solves the possible instability of the theory.} The last solution points decidedly to a version of the Pati-Salam model~\cite{Chamseddine:2015ata}, an avenue also investigated from a phenomenological angle in~\cite{Aydemir:2014ama, Aydemir:2015nfa, Aydemir:2016xtj, Yang:2017vjp, Khozani:2017ykq,  Aydemir:2018cbb}.

 A.\ Devastato, P.\ Martinetti and myself in~~\cite{Devastato:2013oqa, Devastatao:2014xga, Devastato:2014bta} proposed a solution: a \emph{grand} symmetry defining what we termed a \emph{Grand Symmetry} based on \formu{\mathbb M(\mathbb H)_4\oplus \mathbb M(\mathbb C)_8}. In the following I will describe this model in more detail.

In noncommutative geometry the usual grand unified groups, such as \formu{SU(5)} or \formu{SO(10)} do not work, this is discussed in Ex.~\ref{gut}. There are very few representations of associative algebras, as opposed to groups. Finite dimensional algebras only have \emph{one} nontrivial irreducible representations.
 {Fortunately in the standard model there are only weak doublets and colour triplets, so it works}.
 
Recall that a finite ``manifold'' is an algebra of the kind~\eqref{manifoldalgebra} acting on a \formu{2(2a)^2} dimensional Hilbert space. So far we had \formu{a=2, \  2(2a)^2=32 \times 3=96}. The numerology goes well. Let me stress once again how ``lucky'' this is. 

Let us investigate the possibility of enlarging the symmetry by considering values of $a$ in~\eqref{manifoldalgebra} larger than two. For $a=3$ there does not appear to be a realistic model. Let us therefore consider $a=4$, in this case the algebra is
\be
\mathcal A_G=M_{4} (\mathbb H) \oplus M_8(\mathbb C) \label{grandalg}
\ee
one requires a \formu{2(2\cdot 4)^2=128} dimensional space. (384 taking generations into account). This is exactly the dimension of the Hilbert space if we take the fermion quadruplication discussed in Sect.~\ref{SMtriple} into account. This overcounting had been perceived as a nuisance, if not a problem. One had to project states out, and the unphysical redundancy was unexplained. 
Let us now to look at Hilbert space with different eyes
\formula{\mathcal{H}=sp(L^{2}(M))\otimes\mathcal{H}_{F} =L^2(M)\otimes\mathsf H_F}
where now the dimensions of \formu{\mathsf H_F} is 384. We stress htat we have simply unraveled the Lorentz indices of the Dirac spinors. We now have to represent the algebra~$\mathcal A_F$ of~\eqref{grandalg} on $\mathcal A$ maintaining all properties which makes the space a noncommutative manifold. Again, there is warranty this happens, but if is possible to do it.
However, this time the algebra does not act diagonally on the spinor Lorentz indices, on the contrary it mixes them up nontrivially. 

We have to ensure the boundedness of the commutation of the Dirac operator with the element of the algebra, i.e.\ boundedness of the one-form, for a at least dense subalgebra of $\mathcal A_G$. This was automatic when the finite part of algebra $\mathcal A_F$ was acting diagonally on the spinor indices, this is now no longer the case. Therefore one has to to consider a particular form of spectral triple: \emph{twisted} triples~\cite{ConnesMoscovici, Devastato:2014bta, Landi:2017cka, Landi:2016mfh}. This means that instead of the usual commutator, the condition for the manifold spectral triple use a twisted commutator defined by
\be
[D,a]_\rho=Da-\rho(a)D
\ee
with $\rho$ is a inner automorphims of the algebra. This twist has interesting connections with the Wick rotation, as described in~\cite{Devastato:2017rlo}.
We refer to~\cite{Devastato:2013oqa, Devastatao:2014xga, Devastato:2014bta} for all details. The key point is that in the process, spacetime indices, related to the Euclidean symmetries, mix with internal, gauge indices.

We envisage this \emph{Grand Symmetry} to belong to a pre geometric phase of the universe, and that the usual spacetime, and the almost commutative geometry of the standard model, only emerge at lower energies. For this it is necessary that some breaking mechanism be at work. This is possible. At very high energy all of the elements of \formu{D_F}, including the ones relative to the top quark mass, may be considered negligible, What will causes the breaking is the spinor part of the ``free' part of the Dirac operator: \formu{\slashed \del}. Compatibility with the order one condition of this operator (remember that the algebra is not diagonal on the spinor indices, and that we have to use twisted commutators) leads to a phase in which the symmetries of the phase space emerge, and there is  the reduction of the algebra \formu{\mathcal A_{LR}=\mathbb M_2(\mathbb H)\oplus \mathbb M_4(\mathbb C)}, which in turn is broken by the grading to $\mathcal A_{LR}$ of~\eqref{ALR}.

And there is an added, fundamental,  bonus: {this grand algebra, and a corresponding \formu{D} operator, have ``more room'' to operate. Although the Hilbert space is the same, the fact that we abandoned the factorization of the internal indices, gives us more entries to accommodate the Majorana masses.} {Hence we can put a Majorana mass for the neutrino in such a way to obtain a nontrivial commutator,  {and at the same time satisfy the order one condition}. Then the one form corresponding to this \formu{D_\nu} will give us the by now famous field \formu{\sigma}, which can only appear before the transition to the geometric spacetime.}
{The natural scale for this mass is to be above a transition which gives the geometric structure. Therefore it is natural that  it may be at a high scale.}

The grand symmetry is no ordinary gauge symmetry, there is never a gauge \formu{SU(8)} in the game just to mention one difference. {It represents a phase in which the internal noncommutative geometry contains also the spin structure, even the Lorentz (Euclidean) structure of space time in a mixed way.} By not being Lorentz invariant (although there is no preferred direction), the Coleman-Mandula theorem is evaded. {The differentiation between the spin structure of spacetime, and the internal gauge theory comes as a breaking of the symmetry, triggered by \formu{\sigma}, which now appears naturally has having to do with the geometry of spacetime.}{The fermion doubling is turning from being a problem, into a useful resource. In the next section this resource will be further put to good use. }
 
 \section{EUCLID vs.\ LORENTZ \label{EuclidvsLorentz}}

Among the shortcomings of the approach we are discussing there are the issues of being compact and Euclidean. Let us mention that the issue of compactness, and boundary terms, are quite important for the spectral action.  The spectral techniques in a presence of the boundary are discussed for example in~\cite{ConnesChamseddineboundary1, ConnesChamseddineboundary2, Vassilevichboundary, Kurkov:2017cdz, Kurkov:2018pjw}.  
There are various studies which connect spectral triples and Lorentz signatures. Among them let us mention  Krein spaces~\cite{Strohmaier:2001zx,PaschkeSitarz, vandendungen, BrouderBiziBesnard}, covariant approaches \cite{PaschkeVerch}, {Wick rotations on pseudo-Riemanninan structures~\cite{KoenMarioAdam}}, or algebraic characterizations of causal structures~\cite{Moretti, FrancoEckstein, BesnardBizi}.

In this section we will discuss the issue of the signature of the space, mainly from the point of view of the Wick rotation, following~\cite{direstraits}.
The use of Euclidean signature for the action is very common.  The procedure followed is called Wick rotation, a procedure to change the signature of field theory. It consists (loosely speaking) in ``rotating'' the time derivative in the complex plane \formu{t\to it}. This changes the signature of space time from a Lorentzian metric to a Euclidean one. It renders some integrals, which would be oscillatory in the functional integration, convergent since \formu{e^{i t}\to e^{-t}}. In some cases other regularizations work as well, and in principle they are just equivalent procedure which can work always, even if the technical difficulties can be very different.
Then one Wick rotates \emph{back}, i.e., undoes an operation. But in the spectral approach we cannot start unless we have an Euclidean theory. So we are not ``going back'', in going to the Lorentz signature we are venturing in unchartered territory.

{Usually a Wick rotation is indicated as the transformation \formu{t\to\ii t}, even if a more correct procedure would be to rotate the vierbein~\cite{Visser}. Namely for each $F$, which depends on vierbeins}
\formula{
\mbox{Wick:} \quad F\left[e_{\mu}^0, e_{\mu}^j \right]
\longrightarrow \quad F\left[\ii e_{\mu}^0, e_{\mu}^j\right],
\,\, \, j = 1,2,3.]}
The inverse (which is what usually people call Wick rotation) is
\formula{
\mbox{Wick}^*:\quad F\left[e_{\mu}^0, e_{\mu}^j \right]
\longrightarrow \quad F\left[-\ii e_{\mu}^0, e_{\mu}^j\right]
\,\, \, j = 1,2,3. }
For the bosonic part of the spectral action things go relatively without problems, the prescription is clear and the action is rotated into a new one which makes the partition function convergent.
\formula{\mbox{Wick:} \quad S^{\rm E}_{\rm bos}\left[\rm{fields},g_{\mu\nu}^{\rm E}\right]
\longrightarrow S^{\rm E}_{\rm bos}\left[\rm{fields},-g_{\mu\nu}^{\rm M}\right] \equiv -\ii 
S^{\rm M}_{\rm bos}\left[\rm{fields},g_{\mu\nu}^{\rm M}\right]}

{The fermionic sector requires some extra considerations.}
The group \formu{\mathrm{Spin(1,3)}} is quite different from \formu{\mathrm{Spin(4)}}. Indeed \formu{\gamma} matrices, generators, charge conjugation, change. Also the fermionic action changes, since the quadratic forms have to be invariant under the proper group transformations.
\formula{
\bar\psi\, \gamma^{A}_{\rm M}\, e^{\mu}_A \left(
\left[\nabla^{\rm{LC}}_{\mu}\right]^{\rm M} + \ii A_{\mu}\right)
\psi ,
\quad\bar\psi \psi}
{where \formu{\bar\psi \equiv \psi^{\dagger} \gamma^0} and \formu{\nabla_{\mu}^{\rm{LC}}} are the covariant derivative on the spinor bundle with the Levi-Civita spin-connection, which is different for for Lorentzian and Euclidean signature cases.}
The corresponding terms with the required Spin(4) invariance are:
\formula{
\psi^{\dagger}\,\gamma^{A}_{\rm E}\, e^{\mu}_A \left[\nabla^{\rm{LC}}_{\mu}\right]^{\rm E}
\psi,
\quad \psi^{\dagger} \psi } 
 The charge conjugations operators are:
\formula{
C_{\mathrm M}\psi = -\ii\gamma^2_{\mathrm M}  \psi^*\ \ ; \ 
C_{\mathrm E}\psi =\ii\gamma^0_{\mathrm E}\gamma^2_{\mathrm E}=\hat C_{\mathrm E}  \psi^*}

Interestingly enough, the Majorana mass term is the same in both cases:
\formula{\underbrace{\left(C_{\mathrm E}\psi\right)^{\dagger}\psi}_{Spin(4)~\mathrm{inv}} 
= (-i\gamma^0_{\mathrm E}\gamma^2_{\mathrm E}\psi^*)^\dagger\psi
=\overline{(\gamma^2_{\mathrm M}\psi^*)}\psi
=-\underbrace{i\, \overline{\left(C_{\mathrm M}\psi\right)}\psi}_{Spin(1,3)~\mathrm{inv}}}
Also the spacetime grading is the same in the two cases:
\formula{\gamma^5 = \gamma^0_{\rm E}\gamma^1_{\rm E}\gamma^2_{\rm E}\gamma^3_{\rm E}
= \ii \gamma^0_{\rm M}\gamma^1_{\rm M}\gamma^2_{\rm M}\gamma^3_{\rm M}}
so that the definition of left and right spinor is the same.

The difference between \formu{\psi^\dagger} which appers in the Euclidean, and the Lorentzian \formu{\bar\psi}  is the presence of a \formu{\gamma^0} which must be inserted in the Lorentzian case. Consider the fermionic action of~\eqref{SFwithJ}.Thanks to the extra degrees of freedom, the insertion of \formu{\gamma^0} by hand is not needed for this action, which therefore deals with slightly different structures. 
{The fermionic action is build in any case contracting the a conjugate spinor with an operator acting on a spinor. Let us look at the charge conjugation}

We remind that the spacetime part of the Hilbert space splits into eigenspaces of chirality, each of which has two components, for particles and antiparticles
\formula{\mathrm{Sp}(M)=H_\mathcal L\oplus H_\mathcal R }
{with our conventions a the antiparticle of a left particle is right, and viceversa.}
At the same time the internal space has a similar decomposition given by the internal grading \formu{\gamma}
\formula{H_F = H_L \oplus H_R \oplus H_L^c \oplus H_R^c}
{One problem with the quadruplication is the presence of ``mirrors'', states which have different chiralities. They have to be projected out, defining \formu{\mathcal H_+}}
\formula{H_+=(H_L)_{\mathcal L}\oplus(H_R)_{\mathcal R}\oplus(H_L^c)_{\mathcal R}\oplus(H_R^c)_{\mathcal L}= P_+\,H,\quad P_+\equiv\frac{\mathbb 1+\Gamma}2}

This takes care of half of the extra degrees of freedom. The fermionic action is then defined as
\formula{
S_F= \frac{1}{2}\langle J\psi,D_A\psi\rangle \ \ \psi\in H_+ 
}
{
with
\formula{
J= C_{\mathrm E}\otimes J_F }
 and 
\formula{J_F = \left( \begin{array}{cccc}
0 & 0 & \mathbb 1 & 0\\
0 & 0 & 0 & \mathbb 1 \\
\mathbb 1 & 0 & 0 & 0\\
0 & \mathbb 1 & 0 & 0 
\end{array}
\right)\circ cc. }}
{The action  reproduces correctly the Pfaffian i.e.\ the functional integral over fermions, but this procedure only takes care of half of the extra degrees  of freedom. In processes like scattering, after quatization,  it is important to have the correct Hilbert space of incoming and outgoing particles.}

However, in the bosonic spectral action the operator \formu{D} is present, not \formu{DP_+}, which is not Hermitian and not a square root of the Laplacian.
{The extra degrees of freedom are taken care by the Wick rotation. It is in fact necessary to first perform the Wick rotation in order to eliminate the charge conjugation doubling}.
A naive attempt to remove it from the action with the \formu{J}
would break the Euclidean Spin(4) symmetry. {Only the combination of Wick rotation (and identification of states described below) and the projection renders the action viable for physical applications, and free of the fermion doubling}

{Let us see the procedure with some more detail, referring to~\cite{direstraits} for a complete treatment.} {First we rotate the action as in the bosonic case:}
\formula{\mbox{Wick rotation:} \quad - S_F^{\rm E} \left[\mbox{spinors}, e_{\mu}^A \right] 
\longrightarrow  \ii S_F^{\rm M\,doubled}\left[\mbox{spinors}, e_{\mu}^A \right]}
{We now have a  Lorentz invariant  fermionic action  invariant under \formu{Spin(1,3)}   but still exhibiting a doubling.
The spinors are in $H_+$, which is not anymore a Hilbert space with respect to the $Spin(1,3)$ invariant inner product}

{The remaining doubling consists in the presence  of spinors from  all four subspaces of \formu{H_+}: 
\formula{\left(H_L^c\right)_{\mathcal R}, \left(H_R^{c}\right)_{\mathcal L}, \left(H_L\right)_{\mathcal L}, \left(H_R\right)_{\mathcal R}}}
{The physical Lagrangian depends on spinors just from the last two subspaces.}

{After the Wick rotation we should perform  the following identification}
\formula{
\left\{\begin{array}{c}
\left(\psi_L^{c}\right)_{\mathcal R} \in  \underbrace{ \left(H_L^c \right)_{\mathcal R}  }_{\in H_+}
\quad\mbox{identified with}\quad
 C_{\mathrm M}\left(\psi_L\right)_{\mathcal L} , \quad \left(\psi_L\right)_{\mathcal L}\in\underbrace{\left(H_L\right)_{\mathcal L}}_{\in H_+}  \\
\left(\psi_R^{c}\right)_{\mathcal L} \in  \underbrace{ \left(H_R^c \right)_{\mathcal L}  }_{\in H_+}
\quad\mbox{identified with}\quad
 C_{\mathrm M}\left(\psi_R\right)_{\mathcal R} , \quad \left(\psi_R\right)_{\mathcal R}\in\underbrace{\left(H_R\right)_{\mathcal R}}_{\in H_+}
 \end{array} \right. . \label{fWick2}
}
This step leads to the same formula of Barrett~\cite{Barrett}, who considered the spectral action and the standard model starting directly with a Dirac operator and $\gamma$ matrices taken from the Lorentzian signature case. 

We can then apply the procedure to the spectral action. First we restore Lorentz signature in the action:
\bea
 -S_{F}^{\rm E}&\rightarrow& -\int d^4 x\sqrt{-g^{\rm M}} \left[
 \begin{array}{c}
C_{\rm E} \left(\psi_L^c\right)_{\mathcal R} 
\\ C_{\rm E} \left(\psi_R^c\right)_{\mathcal L} 
\end{array}
\right]^{\dagger}\left[
\begin{array}{cc}
\ii \slashed\nabla^{\rm M} & \ii M_D \\
\ii M_D^{\dagger} & \ii\slashed\nabla^{\rm M} 
\end{array}
\right] \left[
 \begin{array}{c}
\left(\psi_L\right)_{\mathcal L} \\ \left( \psi_R \right)_{\mathcal R} 
\end{array}
\right] \label{qq} \nonumber\\
&&-\frac{\ii}{2}\int d^4x \sqrt{-g^{\rm M}}\left\{ \left[C_{\mathrm E}\left(\psi_R\right)_{\mathcal R}\right]^{\dagger}M_M \left(\psi_R\right)_{\mathcal R}
 + \left[C_{\mathrm E}\left(\psi_R^c\right)_{\mathcal L}\right]^{\dagger}M_M^{\dagger} \left(\psi_R^c\right)_{\mathcal L} \right\}
 \eea
 This action is Lorentz invariant under. No 
modification of the inner product, like the insertion of \formu{\gamma^0}, is needed.

Since 
\formula{
C_{\rm E}  = \ii \gamma^0 C_{\rm M}} we have the manifestly Lorentz invariant action:
\bea
S_{F}^{\rm M} &=& \int d^4 x\sqrt{-g^{\rm M}}
\left\{ 
\overline{\left(\psi_{\mathcal L}\right)}\, \ii \slashed\nabla^{\rm M}\psi_{\mathcal L}
+ \overline{\left(\psi_{\mathcal R}\right)}\, \ii \slashed\nabla^{\rm M}\psi_{\mathcal R}\right. \nonumber\\
&&-\left. \left[ \overline{\left(\psi_{\mathcal L}\right)}\,H\,\psi_{\mathcal R} 
+\frac{1}{2} \overline{\left[C_{\rm M}\left(\psi_{\mathcal R}\right)\right]}\,\omega\,\psi_{\mathcal R}
+ \mbox{c.c.}\right] 
\right\}
\eea

%
%
%
We still have extra degrees of freedom since each quantity which carries the index \formu{``c"} is independent from the one which does not.
It is remarkable that the path integral is not sensitive to the charge conjugation doubling, in particular
 the Pfaffian is reproduced correctly since
  \formula{ 
 \int[d \bar \psi][ d\psi] e^{\ii  \int d^4 x\, \bar\psi\,\ii\slashed{\partial}^{\rm M}\psi} = 
 \int[d \bar \xi ][ d\psi] e^{\ii  \int d^4 x\, \bar\xi\,\ii\slashed{\partial}^{\rm M}\psi} \label{dprop}}

The correct identification of the Hilbert space is necessary. The Lorentzian theory has to be \emph{quantized},
and the quantum Hilbert space of asymptotic states has to be constructed.
Such a space is usually referred in physical literature as a ``Fock space". The Hamiltonian coming out of this action is not Hermitian in the Fock space. This is solved with the identification above. The rest is a straightforward exercise. In the end we obtain the correct Lorentzian signature action that you will find in textbooks.

What have we learned? The most intriguing element is that the Euclidean fermionic action, which uses in a crucial way the real structure of the spectral triple, and needs the fermionic quadruplication, is naturally rotated  in the Lorentzian, with the elimination of the extra degrees of freedom.

\section{CONCLUSIONS}

{Noncommutative geometry starts with a view of geometry based on spectral properties, and is geared towards a profound generalization historically opened by the necessity to describe the quantum world}
But then noncommutative point of view grew to become more a framework, or even a \emph{philosophy} for which what is fundamental are not anymore the points, but rather the algebraic structures that we can build over them.

 I tried in the lectures and in these proceeding  to give  the flavour of an application to the physics of fundamental interactions. What we are doing is to understand the noncommutative geometry of the standard model. This view is not the ``party line'' of particle physicists, but nevertheless not only gives a more general framework, which may lead to a more profound understanding, but also makes it conceivable that it may an actual contribution to phenomenology, and confront itself with experiments in a useful way.

There is one aspect of the spectral action, with possible phenomenological applications I did not discusss for lack of spacetime (time during the lectures, space for the proceedings). The spectral action is a curved background gives the usual Einstein-Hilbert action, but it also contains some extra terms proportional to the square of the Riemann tensor. This leads to studies with potential applications to cosmology. A partial list of references on the application of the spectral techniques to cosmology is~\cite{Nelson:2008uy, Marcolli:2009in, Nelson:2009wr, Buck:2010sv, Marcolli:2010fb, Kurkov:2013gma, Lambiase:2015yia}.

\subsection*{Acknowledgments}
I am first of all grateful for the inexhaustible energy of George Zoupanos which propels the Corfu summer institute, and to the opportunity to give the series of lecture in the training school of which these notes are based. 
I also acknowledges the COST action QSPACE which not only organised the school, but also supported my participation. I am further grateful to the INFN Iniziativa Specifica GeoSymQFT, the Spanish MINECO under project MDM-2014-0369 of ICCUB (Unidad de Excelencia `Maria de Maeztu') for support.  

\appendix

\section{EXERCISES AND SOLUTIONS}

{\sc Note: Both the text and the solutions of the exercises were meant as discussion pointers during the lectures and  exercise sessions of the school.  Therefore they necessarily lack the rigour and completeness one would expect in a textbook.}

\begin{itemize}
\item[\ref{ex1}]  \emph{Find the states (and the pure ones) for the algebra of ${n\times n}$ matrices.}

We know the answer from quantum mechanics. Pure states correspond vectors of the Hilbert space on which the matrices act, i.e.\ $n$ dimensional vectors. Nonpure states correspond to density matrices, i.e.\ Hermitean matrices of trace 1 and positive eigenvalues.  

The first statement can be seen as follows. The algebra of matrices is also a Hilbert space with inner product $\Tr b^* a$. By Riesz theorem then any linear functional will be an element of the algebra. Then we set
\be
\phi(a)=\Tr \rho a
 \ \ ; \ \ 
\phi(\mathbb 1)=1 \Rightarrow \Tr\rho=1
\ee

The fact that the matrix can be positive means that it should be Hermitean, and without loss of generality we can choose a basis in which it is diagonal, and being positive with positive eigenvalues. The only way for this diagonal matrix not be be expressible as convex sum of other matrices of this kind is is to have all eigenvalues vanishing except one (which should have value one). This is a pure state, but, considering for example 
\be
\rho=\begin{pmatrix} 1 & 0 &\cdots\\ 0 & 0 & \cdots \\ \vdots & \vdots  & \ddots \end{pmatrix}
\ee
then defining $\varphi=\begin{pmatrix} 1\\ 0 \\ \vdots\end{pmatrix}$ we have
$\phi(a) =\Tr\rho a = \varphi^\dagger a \varphi$.

\item[\ref{2}]  \emph{Prove that ${\mathcal N_\phi}$ is an ideal. } 

This is a one-line proof: it follows from 
 
\be\phi(a^* b^* b a) \leq \|{b}\| ^2 \phi(a^* a)
\ee
 since $\|\phi\|=\sup \{
|\phi(a)| ~|~ \|a\| \leq 1 \}=1$, i.e.\ the norm, which is one, is the supremum over the vectors of norm less than one, and  $\|ab\|\leq\|a\|\, \|b\|$.

\item[\ref{3}] \emph{Perform the GNS construction for ${\mathrm{Mat}(n,\mathbb C)}$ starting from a pure state. }

We will do the construction in gory detail for the two dimensional case, the generalization being straightforward.

Consider the matrix algebra $\mathrm{Mat}(n,\mathbb C)$ with the two
pure states
\be
\phi_1\left( \left[
\begin{array}{cc} a_{11} & a_{12} \\ a_{21} & a_{22}
\end{array}
\right]\right) = a_{11}~,~~~ \phi_2\left( \left[
\begin{array}{cc} a_{11} & a_{12} \\ a_{21} & a_{22}
\end{array}
\right]\right) = a_{22}~.
\ee
The ideals of elements of `vanishing norm' of the states $\phi_1, \phi_2$ are,
respectively,
\be
\mathcal N_1 = \left\{
\left[
\begin{array}{cc} 0 & a_{12} \\ 0 & a_{22}
\end{array}
\right] \right\} ~,~~~~~
\mathcal N_2 = \left\{
\left[
\begin{array}{cc} a_{11} & 0 \\ a_{21} & 0
\end{array}
\right] \right\}~.
\ee
The associated Hilbert spaces are then found to be
\bea
&& {\mathcal H}_1 = \left\{ \left[
\begin{array}{cc} x_1 & 0 \\ x_2 & 0
\end{array}
\right] \right\} ~\simeq~ \mathbb C^2 = \left\{ X = \left(
\begin{array}{c} x_1 \\ x_2
\end{array}
\right) \right\}~, \ \hs{X}{X'} = x_1^* x_1' + x_2^* x_2'\ .  \nonumber \\
&& ~~ \nonumber \\
&& {\mathcal H}_2 = \left\{ \left[
\begin{array}{cc} 0 & y_1 \\ 0 & y_2
\end{array}
\right] \right\} ~\simeq~ \mathbb C^2 = \left\{ X = \left(
\begin{array}{c} y_1 \\ y_2
\end{array}
\right) \right\}~, \ \hs{Y}{Y'} = y_1^* y_1' + y_2^*
y_2'~.\nonumber\\
\end{eqnarray}
As for the action of an element $A \in {\bf M}_2(\mathbb C)$ on
${\mathcal H}_1$ and ${\mathcal H}_2$, we get
\bea
&& \pi_1(A)
\left[
\begin{array}{cc} x_1 & 0 \\ x_2 & 0
\end{array}
\right] =
\left[
\begin{array}{cc} a_{11}x_1 + a_{12}x_2 & 0 \\
a_{21}x_1 + a_{22}x_2 & 0
\end{array} \right]
\equiv A \left(
\begin{array}{c}
x_1  \\
x_2
\end{array}
\right) ~, \nonumber \\
&& \pi_2(A)
\left[
\begin{array}{cc} 0 & y_1 \\ 0 & y_2
\end{array}
\right] =
\left[
\begin{array}{cc} 0 & a_{11}y_1 + a_{12}y_2 \\
0 & a_{21}y_1 + a_{22}y_2
\end{array} \right]
\equiv A \left(
\begin{array}{c}
y_1 \\
y_2
\end{array}
\right) ~. \label{repr}
\end{eqnarray}
The equivalence of the two representations is provided by the off-diagonal
matrix
\be
U = \left[
\begin{array}{cc}
0 & 1 \\
1 & 0
\end{array}
\right]~,
\ee
which interchanges $1$ and $2$~: $U \xi_1 = \xi_2$. Using the fact
that for an irreducible representation any nonvanishing vector is
cyclic, from (\ref{repr}) we see that the two representations can
be identified.

The procedure generalises to arbitrary $n$, and also, with little work,  to compact operators.

\item[\ref{4}] \emph{Given the algebra of continuous functions on the line ${\mathcal C_0(\mathbb R)}$, consider the two states $\delta_{x_0}(a)=a(x_0)$ and ${\phi(a)=\frac1{\sqrt{\pi}}\int_{-\infty}^\infty dx\, e^{-x^2} a(x)}$. Find the Hilbert space in the two cases. }

In the first case $\mathcal N_\delta$ is composed of all functions which vanish at $x_0$. Two function belong to the same equivalence class of $\mathcal C_0(\mathbb R)/\mathcal N_\delta$ if they differ by any function which vanishes at $x_0$. Therefore they must have the same value in $x_0$, and we can identify the class of equivalence with this value. Therefore the quotient is simply $\mathbb C$. Note that by the same token the vaue of the function at a point is also an irreducible representation of the algebra. Since the algebra is commutative the only IRR are indeed $\mathbb C$.

In the second case $\mathcal N=\emptyset$ since $\phi(a^*a)=\frac1{\sqrt{\pi}}\int_{-\infty}^\infty dx\, e^{-x^2} |a(x)|^2$ cannot be zero if $a\neq 0$. We then have a faithful (but reducible) representation. 

This state is clearly not pure, for example:
\be
\phi(a)=\frac12\phi_1(a)+\frac12\phi_2(a)=\frac1{\sqrt{\pi}}\int_{-\infty}^0dx\, e^{-x^2} a(x)+\frac1{\sqrt{\pi}}\int_{0}^\infty dx\, e^{-x^2} a(x) 
\ee

\item[\ref{5}] \emph{Make ${\mathcal A^N}$ into a Hilbert module, and discuss its automorphisms in the cases ${\mathcal A}$ is ${\mathrm{Mat}(n,\mathbb C)}$ or ${C_0(M)}$.}

An element of $A\in\mathcal A_N$ is just an $N$-ple of element of the algebra $\{A_i\}$ and I can define a right (left) module as
\be
(a A)_i=aA_i\ ; \ (Aa)_i=A_ia
\ee
The inner product can be easily defined 
\be
\langle a,b\rangle_{\mathcal A}=\sum_i a_i^\dagger b_i
\ee
The
corresponding norm is
\be
\norm{(a_1, \cdots, a_N)}_\mathcal A := \norm{\sum_{k=1}^n a_k^* a_k}~.
\ee
That $\mathcal A^N$ is complete in this norm is a consequence of the
completeness with respect to its norm. Parallel to the
situation of the previous example, when the algebra is
unital, one finds that $\mathrm{End}_\mathcal A(\mathcal A^N) \simeq \mathrm{End}^0_\mathcal A(\mathcal A^N)
\simeq {\mathbb M}_n(\mathcal A)$, acting on the left on $\mathcal A^N$. The
isometric isomorphism $\mathrm{End}^0_\mathcal A(\mathcal A^N) \simeq {\mathbb
M}_n(\mathcal A)$ is now given by
\begin{eqnarray}
&& \mathrm{End}^0_\mathcal A(\mathcal A) \ni \ket{(a_1, \cdots, a_N)} \bra{(b_1, \cdots,
b_N)} ~\mapsto~ \left(
\begin{array}{ccc}
a_1 b^*_1 & \cdots & a_1 b^*_N \\
\vdots & ~ & \vdots \\
a_N b^*_1 & \cdots & a_N b^*_N \\
\end{array}
\right)~, \nonumber \\
&&\forall ~a_k, b_k \in \mathcal A~,
\end{eqnarray}
which is then extended by linearity.

The automorphisms of this module are simple $N\times N$ matrices with entries in the element of the algebra. If $\mathcal A$ is the algebra 
Note that for $\mathcal A=\mathrm{Mat}(n,\mathbb C)$ the same space is also a Hilbert \emph{space}, i.e.\ a Hilbert module over $\mathbb C$
\be\langle a,b\rangle_{\mathbb C}=\sum_i \Tr a_i^\dagger b_i
\ee
While for $\mathcal A=C_0(M)$ $$\langle a,b\rangle_{\mathbb C}=\sum_i \int \dd\mu \,  a_i^\dagger b_i
$$
and the completion of the integral gives $L^2$.

\item[\ref{6}]\emph{The algebras ${\mathbb C, \mathrm{Mat}(\mathbb C,n)}$ and and ${\mathcal K}$, compact operators on a Hilbert space are all Morita equivalent. Find the respective bimodules.}

\def\lhs#1#2#3{{_{#3}\!}\left\langle #1,#2\right\rangle}
\def\rhs#1#2#3{\left\langle #1,#2\right\rangle\!{_{#3}}}

\def\lhsa#1#2{{\ca\!}\left\langle #1,#2\right\rangle}
\def\rhsa#1#2{\left\langle #1,#2\right\rangle\!{_{\ca}}}

For any integer $n$ the algebras ${\mathbb M}_n(\mathbb C)$ and $\mathbb C$
are Morita equivalent and the equivalence ${\mathbb
M}_n(\mathbb C)$-$\mathbb C$ bimodule is just $\mathcal E=\mathbb C^n$. The
left action of ${\mathbb M}_n(\mathbb C)$ on $\mathcal E$ is the usual
matrix action on a vector, while $\mathbb C$ acts on the right on
each component of the vector. The hermitian products for any two
vectors $u=(u_i)$ and $v=(v_i)$ are $\hs{v}{w}_{\mathbb C} :=
\sum_{i=1}^n\bar{v}_i {w}_i$ and $\lhs{u}{v}{{\mathbb M}_n(\mathbb C)} := \ket{u}\bra{v}$ which reads $\bar
u_i v_j$ in components. It is immediate to verify relation~\eqref{compmorita}:
\be
\lhs{v}{w}{{\mathbb M}_n(\mathbb C)}u=\ket{v}\bra{w}\, {u}= v
\hs{w}{u}_\mathbb C \ .
\ee
In components all of the above expressions are simply $\sum_j \bar
v_iw_ju_j$. The module $\mathcal E=\mathbb C^n$ is free as a right
$\mathbb C$-module while it is projective (of finite type) as a
left ${\mathbb M}_n(\mathbb C)$-module and $\mathbb C^n = {\mathbb
M}_n(\mathbb C) p$ where $p=\ket{v}\bra{v}$ is any rank-one
projection.

Generalizing the procedure one shows that the algebra ${\mathcal K}({\mathcal H})$ of compact operators on a separable Hilbert space
${\mathcal H}$ is Morita equivalent to the algebra $\mathbb C$ with
${\mathcal H}$ the equivalence ${\mathcal K}({\mathcal H})$-$\mathbb C$
bimodule and hermitian products $\lhs{v}{w}{{\mathcal K}({\mathcal H})} :=
\ket{v}\bra{w}$ and $\hs{v}{w}_{\mathbb C} := \hs{v}{w}_{{\mathcal H}}$. Again ${\mathcal H} = {\mathcal K}({\mathcal H}) p$ where
$p=\ket{v}\bra{v}$ is any rank-one projection but now ${\mathcal H}$
is not finite generated (i.e.\ finite dimensional) over
$\mathbb C$.

If $M$ is a locally compact Hausdorff topological space, then from
the previous considerations follows that for any integer $n$ the
algebra ${\mathbb M}_n(\mathbb C) \otimes C_0(M) \simeq {\mathbb
M}_n(C_0(M))$ is Morita equivalent to the algebra $C_0(M)$.

\item[\ref{7}] \emph{Take ${\mathcal A=\mathcal C(\mathbb R)}$ and  ${D=i\del_x}$. Prove that the distance among pure states gives the usual distance among points of the line ${d(x_1,x_2)=|x_1-x_2|}$ .}

Given a function $a(x)$ of the algebra we should consider the supremum of $|a(x_1)-a(x_2)|$, subject to the condition on the norm of $[D,a]\leq 1$ is 
\be
\|[D,a]\|=\|\sup_x \|a'(x)\|
\ee
clearly the supremum is attained for any function which ha $|a'(x)=1|$ in the interval $[x_1,x_2]$, and the behaviour of the function outside is irrelevant (it must be just  such that it eventually goes to zero at infinity). 

In more dimensions, with generic $\gamma$'s with $\{\gamma^\mu,\gamma^\nu\}=g^{\mu\nu}$ the calculation is more involved for $g$ generic, it goes through a Lipschitz norm, and in the end one obtains the usual geodesic distance given by the metric tensor.

\item[\ref{8}] \emph{Find at least two good reasons for which the construction of last lecture cannot be performed (without changes) for a noncompact and Minkowkian spacetime.}

On a noncompact space a derivative operator such as the original Dirac operator does not have a discrete spectrum, and therefore our construction based on the eigenvalues of $D$ will not count. This a technical infrared problem, and on can circumvent it putting the system ``in a box''. A Minkowkian compact space in somewhat problematic as well, and not immediately compatible with causality, what happens to the light cones? We will see in the course of the lectures that something can be done to create some Minkowskian noncommutative geometry.

\item[\ref{9}] \emph{Consider the toy model for which $\mathcal A$ is $\mathbb C^2$ represented as diagonal matrices on $\mathcal H=\mathbb C^m\oplus \mathbb C^n$, and ${D_0=\begin{pmatrix} 0 & M\\M^\dagger &0\end{pmatrix}}$. Find the fluctuated Dirac operator.}

The exercise really has to do with finding one forms. We will solve it using some more sophisticated mathematics, which will help us introduce some concepts we overlooked in the text.

Consider a space made of two points $Y =\{1, 2\}$. The algebra
$\mathcal A$ of continuous functions is the direct sum $\mathcal A = \mathbb C \oplus
\mathbb C$ and any element $f\in\mathcal A$ is a pair of complex numbers $(f_1,
f_2)$, with $f_i = f(i)$ the value of $f$ at the point $i$.
An even spectral triple $(\mathcal A, \mathcal H, D, \gamma)$ is
constructed as follows. The finite dimensional Hilbert space $\mathcal H$
is a direct sum $\mathcal H = \mathcal H_1 \oplus \mathcal H_2$ and elements of $\mathcal A$
act as diagonal matrices
\be
\mathcal A \ni f \mapsto \left(
\begin{array}{cc}
f_1 \mathbb 1_{\dim \mathcal H_1} & 0 \\
0 & f_2 \mathbb 1_{\dim \mathcal H_2}
\end{array}
\right) 
\ee

There is a natural grading operator $\gamma$ given by
\be
\gamma = \left(
\begin{array}{cc}
\mathbb 1_{\dim \mathcal H_1} & 0 \\
0 & - \mathbb 1_{\dim \mathcal H_2}
\end{array}
\right)~.
\ee
An operator $D$ -- being required to anticommute with $\gamma$ -- must
be an off-diagonal matrix,
\be
D = \left(
\begin{array}{cc}
0 & M \\
M^* & 0
\end{array}
\right)~, ~~ M \in {\rm Lin}(\mathcal H_2,\mathcal H_1)~.
\ee
With $f\in\mathcal A$, one finds for the commutator
\be
[D, f] = (f_2-f_1) \left(
\begin{array}{cc}
0 & M \\
-M^* & 0
\end{array}
\right)~,
\ee
and for its norm, $\norm{[D, f]} = |f_2-f_1| \lambda$ with
$\lambda$ the largest eigenvalue of the matrix $|M| =
\sqrt{M^*M}$. Therefore, the noncommutative distance between the
two points of the space is found to be
\be
d(1,2) = \sup\{|f_2-f_1| ~|~ \norm{[D, f]} ~\leq 1 \} =
\frac{1}{\lambda}~.
\ee
Since $D$ is a finite hermitian matrix, this geometry is just `$0$-dimensional' and the only available trace is ordinary matrix trace.

It turns out that for this space, and for
more general discrete spaces as well, it is not possible to
introduce a real structure which fulfils all the requirements. It seems that it is not possible to satisfy
the first order condition.

We now construct the exterior algebra on this two-point space.
The space $\Omega^1\mathcal A$ of universal
$1$-forms can be identified with the space of functions on $Y
\times Y$ which vanish on the diagonal. Since the complement of
the diagonal in $Y \times Y$ is made of two points, namely the
pairs $(1,2)$ and $(2,1)$, the space $\Omega^1\mathcal A$ is two-dimensional
and a basis is constructed as follows. Consider the function $e$
defined by $e(1) = 1, e(2) = 0$; clearly, $(1-e)(1) = 0,
~~(1-e)(2) = 1$. A possible basis for the $1$-forms is then given
by\footnote{Here we are oversimplifying a construction based on cyclic cohomolgy. For details se~\cite{landibook,ticos}}
\be
e\delta e ~, ~~~(1-e) \delta (1-e)~,
\ee
and their values are
\begin{eqnarray}
&& (e\delta e) (1,2) = -1~, ~~~((1-e) \delta (1-e))(1,2) = 0 \nonumber \\
&& (e\delta e) (2,1) = 0~, ~~~~~ ((1-e) \delta (1-e))(2,1) = -1~.
\end{eqnarray}
where I defined $\delta e\equiv [D,e]$.

Any $1$-form  can be written as
$\alpha = \lambda e\delta e + \mu (1-e) \delta (1-e)$, with
$\lambda, \mu \in \mathbb C$. One immediately finds that
\begin{align}
 e [D, e] &= \left(
\begin{array}{cc}
0 & - M \\
0 & 0
\end{array}
\right)~, \\
  (1-e)[D, 1-e] &= \left(
\begin{array}{cc}
0 & 0\\
-M^* & 0
\end{array}
\right)~,
\end{align}
and a generic $1$-form $\alpha = \lambda e\delta e + \mu (1-e) \delta
(1-e)$ is \be
\alpha = - \left(
\begin{array}{cc}
0 & \lambda M \\
\mu M^* & 0
\end{array}
\right)~.
\ee
Hermitean one forms have $\mu=\bar\lambda$.

\item[\ref{10}] \emph{Consider the case for which $\mathcal A$ is the algebra of two copies of function on a manifold, ${C_0(M)\times \mathbb Z_2= C_0(M)\oplus C_0(M)}$, again  represented as diagonal matrices on ${\mathcal H=L^2(M)\oplus L^2(M)}$, and ${D=i \slashed \del \otimes \mathbb 1+\gamma^5\oplus D_F}$. Find the fluctuated Dirac operator. }

The geometry in this case is almost commutative, i.e.\ the product of the ordinary commutative manifold times a finite dimensional space composed by two points. Note that the geometry is still commutative, the topological space is comprised of two copies of $M$.  Accordingly with the considerations of the previous example $\gamma$ and $D_f$ are like the earlier $D$ before, with the matrix $M$ being just a complex number. 

One forms split int two as well. The one due to $\slashed\del$ gives a Hermitean element of the algebra, which corresponds to the potential of a $U(1)\times U(1)$ theory, since the two elements on the diagonal matrix are different.

The part of the potential coming from $D_F$ is instead a \emph{scalar field} which connects the two sheets.
\be
D_A=(D+\slashed A)\otimes\mathbb 1 + \gamma^5 \otimes \begin{pmatrix} 0 & \phi\\ \phi & 0 \end{pmatrix}
\ee
Note that the algebra is made of two diagonal elements, which are eigenvalues of $\gamma$ with $\pm 1$ eigenvalues, hence, heuristically speaking, they are ``right'' and ``left'' sheet. The field $\phi$ connect the two sheets, which is what the Higgs field actually does. This analogy becomes more precise with the spectral action.

\item[\ref{gut}] \emph{Perform a similar construction for a Grand Unified Theory}

The construction will not work for the majority of GUT's. The reason is that an algebra has less irreducible representations than a group. 

The reason for this is simple, an algebra has to accommodate two operations, sum and product, while a group only has one.

The relevant case is $\mathrm{Mat}(n,\mathbb C)$. As algebra it has only one non trivial irreducible representation, the definitory one, i.e.\ $n\times n$ complex matrices. The unitary elements of the algebra form always a group, and in this case the group is $U(n)$, which has is known has an infinity of irreducible representations. If we take a $m$ dimensional representation of $U(m)$ and start allowing also the matrices which we obtain using the sum besides the product then we unavoidably obtain $\mathrm{Mat}(m,\mathbb C)$ or $\mathrm{Mat}(m,\mathbb R)$ according to the representation, but not $\mathrm{Mat}(n,\mathbb C)$.

Partial exception are conplex numbars (one by one matrices)  two by two matrices. For the former we have a representation for any real number, but the algebra  one obtains is of course always $\mathbb C$. The two by two case
is connected with the peculiar characteristics of Pauli matrices. In this case we have the possibility of using quaternions, whose unitary elements are $SU(2)$.

The construction with a gauge group (also made of different factors) requires fermions to transform either in the fundamental representation of the gauge group (or each of its factors) or the trivial one. This is what happens for the standard model, where quarks tranform under the the fundamental representation of $SU(3)$ and leptons under the trivial one, left particles are doublets of $SU(2)$ and right particles are singlets (transform under the trivial representation).

GUT's like $SU(5)$ require the fermions to be represented in a five dimensional spae (and this is OK) as well as 10 dimensional space, and this creates the problem. Likequse for $SO(10)$ which requires a 16 dimensional representation.

An exception is the Pati-Salam model based on $SU4)\times SU(2)\times SU(2)$. In this case lepton number becomes a ``fourth colour'' and right particles transform under the second $SU(2)$.

\end{itemize}

\end{document}